\documentclass[pra,nofootinbib,english,twocolumn,amsmath,amssymb,citeautoscript,showpacs,longbibliography,superscriptaddress]{revtex4-1}

\usepackage{graphicx,color} \usepackage{braket}
\usepackage[colorlinks,urlcolor=blue,citecolor=blue]{hyperref}
\usepackage{mathptmx}
\usepackage[T1]{fontenc}
\usepackage[latin9]{inputenc}
\usepackage{babel}

\newcommand{\Rcal}{\mathcal{R}} \newcommand{\Qcal}{\mathcal{Q}}

\newcommand{\LiK}{${}^{6}\mathrm{Li}$-${}^{40}\mathrm{K}$ }
\newcommand{\LiSix}{${}^{6}\mathrm{Li}$ }
\newcommand{\KForty}{${}^{40}\mathrm{K}$ }

\begin{document}

\title{Multiple phase separation in one-dimensional mixtures of mass- and\\ population-imbalanced attractive Fermi gases}

\author{Sangwoo S.~\surname{Chung}} \affiliation{Department of Physics, University of Cincinnati, Cincinnati, Ohio 45221-0011, USA} \affiliation{Institute for Theoretical Physics Amsterdam and Delta Institute for Theoretical Physics, University of Amsterdam, Science Park 904, 1098 XH Amsterdam, The Netherlands}

\author{C.~J.~\surname{Bolech}} \affiliation{Department of Physics, University of Cincinnati, Cincinnati, Ohio 45221-0011, USA}

\begin{abstract}
We study the attractive Fermi mixture of a \LiK gas in one dimension using the continuous matrix product states variational
ansatz and obtain the $T=0$ phase diagram. We predict an axial density profile that contains four distinct phases trapped induced along one-dimensional (1D) tubes, which is more intricate than those observed in 1D mass-balanced systems or in higher-dimensional gas clouds.  The parameter regimes explored are realistic in view of possible future experiments. This an application of continuous matrix product states to a nonintegrable fermionic system.
\end{abstract}

\maketitle

\section{Introduction}

Two-spin mixtures of degenerate atomic Fermi gases \cite{Giorgini2008,Guan2013,Gubbels2013} have been of extraordinary interest during the past decade, to both atomic and condensed matter physicists, exhibiting a number of remarkable physical phenomena such as the crossover from a Bardeen-Cooper-Schrieffer (BCS) superfluid to a Bose-Einstein condensate (BEC) \cite{Zwierlein2004,Regal2004,Bartenstein2004,Kinast2004,Partridge2005}, fermionic superfluidity \cite{Zwierlein2005,Zwierlein2006,Partridge2006,Partridge2006b}, and exotic pairing \cite{Liao2010}.  In regard to exotic pairing, a one-dimensional (1D) system of ultracold spin-imbalanced $(N_{\uparrow}\ne N_{\downarrow})$ Fermi mixtures has been noteworthy, as a large portion of its phase diagram is predicted \cite{Orso2007,Hu2007} to be in the Fulde-Ferrell-Larkin-Ovchinnikov (FFLO) superfluid phase \cite{Fulde1964,Larkin1965,Casalbuoni2004} where paired particles have unequal chemical potentials.  In response to the theoretical predictions, Liao \emph{et al.} \cite{Liao2010} have created a two-spin mixture of ultracold \LiSix gas, confined in 1D tubes arranged in a 2D optical lattice.  The agreement between their data and the theoretical \cite{Feiguin2007,Casula2008,Kakashvili2009} density profiles suggests that their quasi-1D system is a close representation of the true 1D system \cite{Sun2013} that realizes a spin-imbalanced FFLO-like state.  While the FFLO phase has yet to be confirmed experimentally with certainty, it can potentially be a general pairing phenomenon, found not only in ultracold atomic systems, but in condensed matter \cite{Bianchi2003,Uji2006,Mayaffre2014}, nuclear physics \cite{Stein2014}, and color superconductivity in dense quark matter \cite{Alford2008,Sedrakian2009,Huang2010} as well.

In addition to spin imbalance, a mass imbalance $(m_{\uparrow}\ne m_{\downarrow})$ in the system can be another source for the Fermi surface mismatch between the two species that can lead to FFLO pairing.  Such systems naturally occur, for instance, in quantum chromodynamics, where quarks having different masses can bind together, or in a neutron-proton condensate in nuclear physics.  In ultracold atomic systems, the \LiSix and \KForty gases are particularly well controlled by the experimentalists \cite{Wille2008,Voigt2009,Jag2014,Trenkwalder2014}, lending the \LiK mixture as one of the preferred model systems for theoretical investigations \cite{Gubbels2009,Baarsma2010,Baarsma2013,Wang2013,Pahl2014} of mass-imbalanced Fermi gases.

In this paper we present the zero-temperature phase diagram for the attractive \LiK 1D gas system and in turn make an intriguing prediction of a trap-induced multiple phase separation in the density profile, which can be realized with an ultracold \LiK mixture confined in 1D tubes.  The phases include two oppositely polarized FFLO-like superfluids, which is not expected for equal-mass systems in one or higher dimensions.

Our numerical tool is the continuous matrix product state (cMPS) variational ansatz \cite{Verstraete2010, Haegeman2013a}, which is a recent development of the continuum analog of the matrix product state (MPS) ansatz \cite{Affleck1987,Ostlund1995,Schollwock2011}.  The MPS is equivalent to the density-matrix renormalization group (DMRG) algorithm \cite{White1992,Schollwock2005}, which is arguably the most powerful numerical technique to date for simulating quantum lattice systems in low dimensions.  The cMPS ansatz has demonstrated its capability in predicting ground-state properties of 1D continuum systems of interacting bosons \cite{Verstraete2010,Rincon2015}, Luttinger liquids \cite{Quijandria2014, Quijandria2015}, relativistic fermions \cite{Haegeman2010}, spin-imbalanced fermions \cite{Chung2015}, and excitation properties of bosons \cite{Draxler2013}.  It has also proved useful in the study of fractional quantum Hall states \cite{Zaletel2012}.  While there have been previous studies on mass-imbalanced Fermi gas mixtures in one dimension using bosonization \cite{Cazalilla2005}, finite-temperature studies with mean-field approximation \cite{Parish2007,Roscher2014}, exact diagonalization \cite{Pecak2016}, time-evolving block decimation (TEBD) \cite{Wang2009}, and DMRG \cite{Orso2010,Dalmonte2012}, the cMPS gives a direct means to study 1D continuum systems without resorting to linearization of the spectrum or discretization of the space to apply numerical techniques that were intrinsically tailored for lattice systems \cite{Stoudenmire2012,Dolfi2012}.

\section{Model and Results}

\subsection{Model Hamiltonian and $T=0$ Phase Diagram}

In order to investigate the ground-state properties of the \LiK gas system in one dimension, we consider the Gaudin-Yang Hamiltonian for a $\delta$-function interacting spin-1/2 Fermi gas in a second-quantized form
\begin{equation}
\hat{H}=\int_{0}^{L}dx\sum_{\sigma=\uparrow,\downarrow}\left(\frac{\hbar^{2}}{2m_{\sigma}}\partial_{x}\hat{\psi}_{\sigma}^{\dagger}\partial_{x}\hat{\psi}_{\sigma}-2\Delta\hat{\psi}_{\sigma}^{\dagger}\hat{\psi}_{\bar{\sigma}}^{\dagger}\hat{\psi}_{\bar{\sigma}}\hat{\psi}_{\sigma}\right),
\end{equation}
where $\{\hat{\psi_{\sigma}}(x),\hat{\psi}^{\dagger}_{\sigma'}(x')\}=\delta_{\sigma\sigma'}\delta(x-x')$.  The two fermionic species have masses $m_{\uparrow}$ and $m_{\downarrow}$, and $L$ is the system length.  This system is known to be integrable using the Bethe ansatz technique for the mass-balanced case only.  We use the pseudo-spin convention where $\uparrow$ and $\downarrow$ designate the \LiSix and \KForty atoms, respectively.  The $\bar{\sigma}$'s denotes the conjugate spin to $\sigma$, and the interaction strength is $\Delta>0$ for an attractive interaction between the two species. We use units where $\hbar = e = k_{q} = 1$, for the Planck constant, the electron charge, and the Coulomb constant, respectively.  Our mass unit is $m_0 = 480/23 \:\mathrm{amu}$.  In these units, $2m_{\uparrow}=23/40$ and $2m_{\downarrow}=23/6$, and the reduced mass $m_{r}=m_{\uparrow}m_{\downarrow}/(m_{\uparrow}+m_{\downarrow})$ and the bound state energy are the same as in the mass-balanced system $(2m_{\uparrow}=2m_{\downarrow}=1)$, namely, $m_{r}=0.25$ and $\epsilon_{b} = 8m_{r}{\Delta}^2/{\hbar}^2 = 2\Delta^2$.

A phase diagram for the attractive \LiK 1D gas can be obtained from minimizing the average free-energy density $f=\langle\hat{F}\rangle/L$ of the system at $T=0$, where
\begin{equation}
\hat{F}=\hat{H}-\int_0^L dx
\left\{\mu\left[\hat{n}_{\uparrow}(x)+\hat{n}_{\downarrow}(x)\right]+h\left[\hat{n}_{\uparrow}(x)-\hat{n}_{\downarrow}(x)\right]\right\},\label{eq:freeenergy}
\end{equation}
with $\hat{n}_{\sigma}(x)=\hat{\psi}_{\sigma}^\dagger(x) \hat{\psi}_{\sigma}(x)$, $\mu=\left(\mu_{\uparrow}+\mu_{\downarrow}\right)/2$, and $h=\left(\mu_{\uparrow}-\mu_{\downarrow}\right)/2$.  
\begin{figure}
\includegraphics[width=\columnwidth]{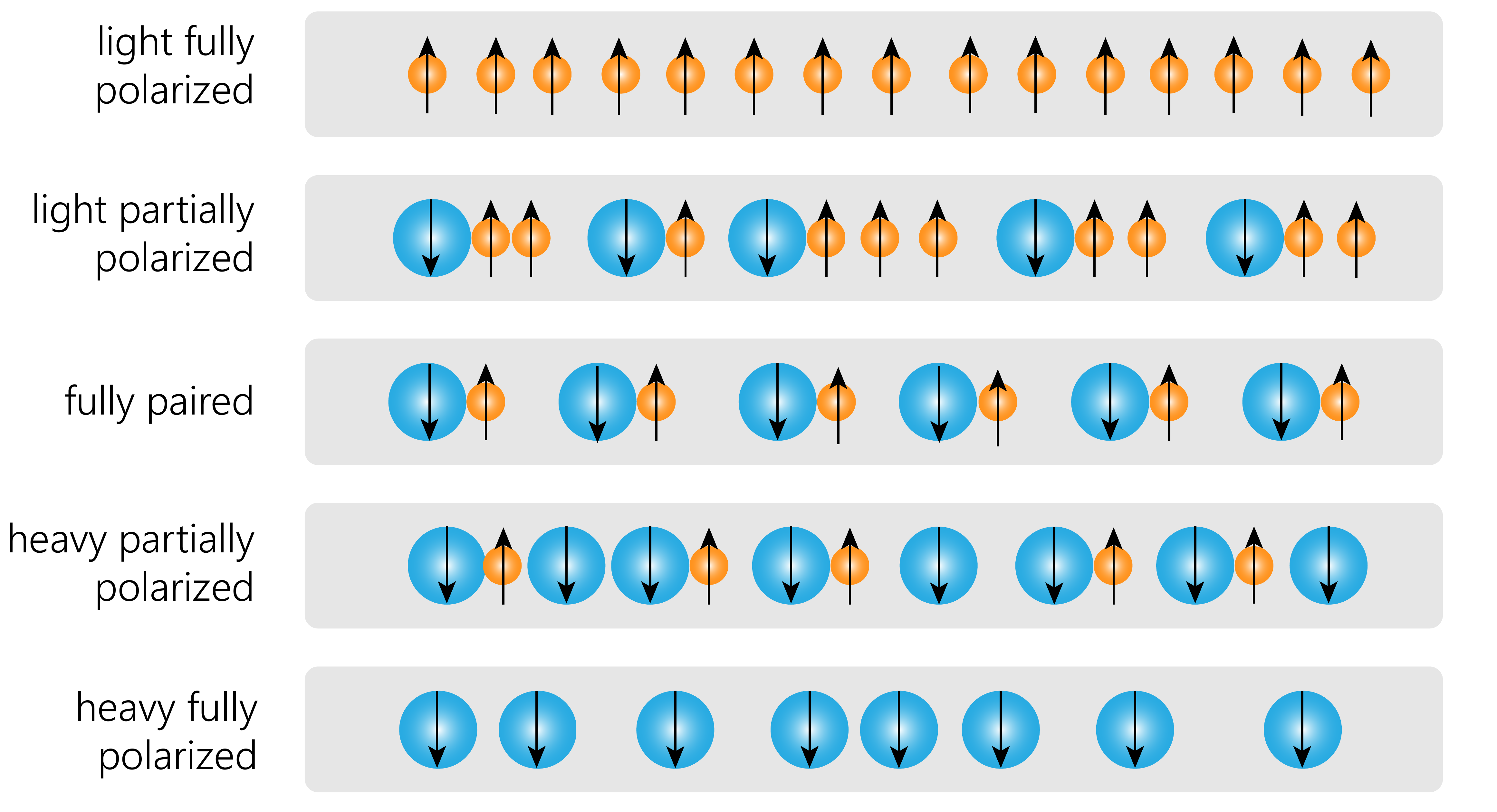}
\caption{\label{fig:phases} The five distinct phases in the one-dimensional attractive \LiK gas system.}
\end{figure}
Besides the vacuum (i.e. $N_{\uparrow}=N_{\downarrow}=0$), there are five distinct phases (Fig.~\ref{fig:phases}):
(i) a light fully polarized state when $N_{\uparrow}>N_{\downarrow}=0$,  (ii) a light partially polarized state when $N_{\uparrow}>N_{\downarrow}>0$, (iii) a fully paired state when $N_{\uparrow}=N_{\downarrow}>0$, (iv) a heavy partially polarized state when $N_{\downarrow}>N_{\uparrow}>0$, and (v) a heavy fully polarized state when $N_{\downarrow}>N_{\uparrow}=0$.

\begin{figure}
\includegraphics[width=\columnwidth]{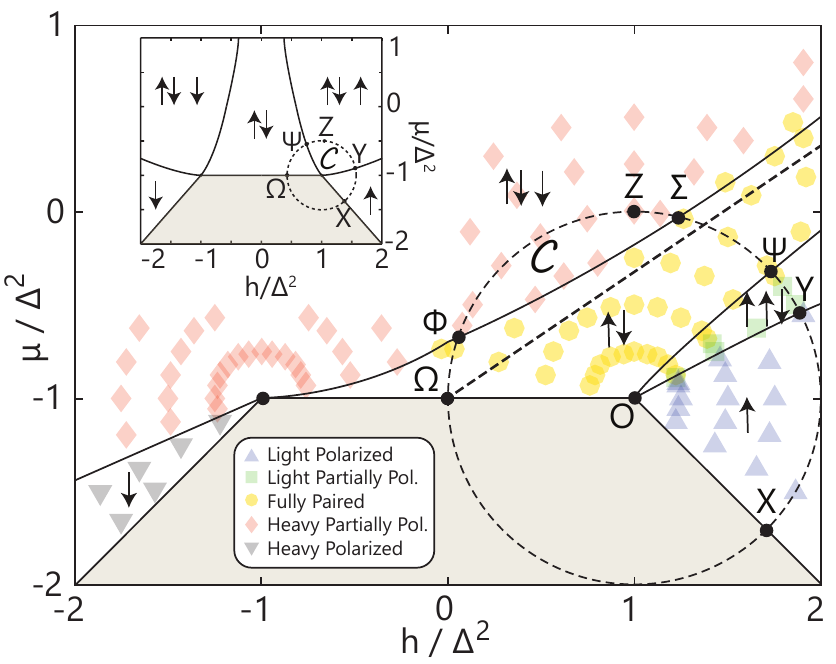}
\caption{\label{fig:pd} The $T=0$ phase diagram for an attractive \LiK gas mixture in one dimension, obtained from the cMPS with a matrix dimension of $D=16$.  The five distinct symbols and colors indicate the distinct ground-state phases as determined from the cMPS and the shaded area indicates the vacuum phase.  The dashed circle (centered at point O) indicates the contour where the data points in Fig.~\ref{fig:energy} have been sampled from.  The points X, Y, $\Psi$, $\Sigma$, $\Phi$, and $\Omega$ along the dashed circle $\mathcal{C}$ denote its intersections with the phase boundaries.  The phase boundaries are estimates and were drawn as a guide to the eye only.  The inset shows the $T=0$ ground-state phase diagram for a mass-balanced system (adapted from Ref.~\cite{Orso2007}).}
\end{figure}

Figure \ref{fig:pd} shows the ground-state phase diagram as determined numerically using cMPSs.  The dashed line that originates from point $\Omega$ indicates the $\mu$ and $h$ values where the Fermi points for the two species coincide in a noninteracting system, shifted vertically down by $\Delta^2$ for the attractive system.  Along this line, $\mu=\frac12\left(\frac{1+m_{\uparrow}/m_{\downarrow}}{1-m_{\uparrow}/m_{\downarrow}}\right)h-\Delta^2$, we expect a BCS type of conventional superfluid, and the line indeed penetrates through the fully paired region.

In the mass-balanced system (Fig.~\ref{fig:pd}, inset), the $\uparrow$ and the $\downarrow$ phases are symmetric about $h=0$.  For the mass-imbalanced \LiK system, on the other hand, in addition to the lack of symmetry about $h=0$, the heavy partially polarized phase occupies a large portion of the phase diagram and extends deep into the positive-$h$ region, whereas the light partially polarized phase is narrow and much smaller.  Within a local-density approximation, decreasing $\mu$ at a fixed $h$ is equivalent to moving from the center of the harmonic trap to its edge.  For $h>\Delta^2$, one could create a sufficiently deep harmonic potential that encompasses four distinct phases, a heavy partially polarized core, followed by fully paired, light partially polarized and light-polarized shells, as discussed in Sec.~\ref{section:PS}.

\begin{figure}
\includegraphics[width=\columnwidth]{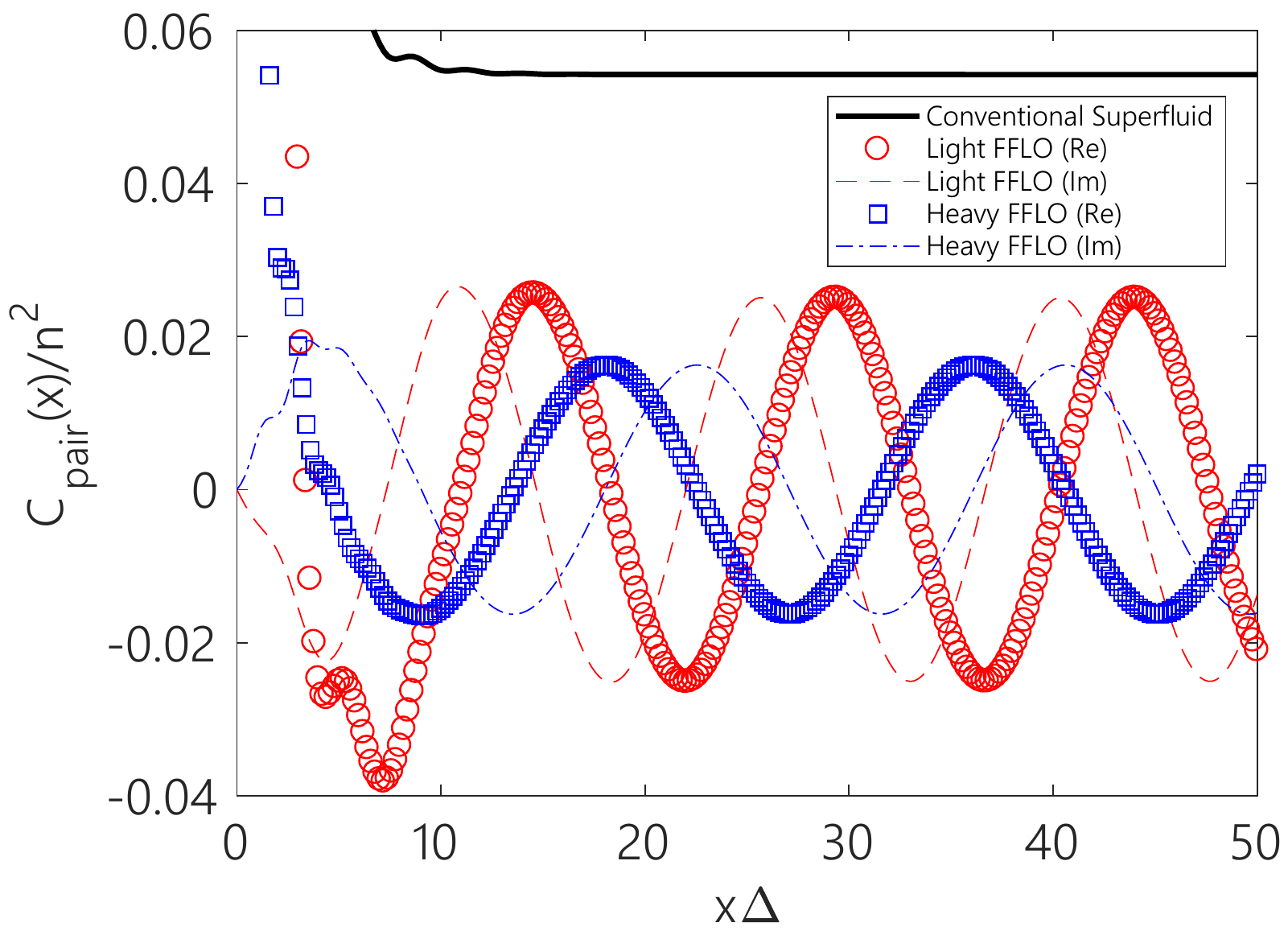}
\caption{\label{fig:cpair} Pair correlation function $C_{\mathrm{pair}}(x)\equiv\langle \psi_{\uparrow}^{\dagger}(0)\psi_{\downarrow}^{\dagger}(0)\psi_{\downarrow}(x)$ $\psi_{\uparrow}(x)\rangle$ for conventional-superfluid, light-FFLO and heavy-FFLO phases, obtained from $(\mu/\Delta^2,h/\Delta^2)$ values (1.707,-0.2929), (1.866,-0.5), and (1,0), respectively.  For the FFLO correlations, the circles and squares indicate the real parts of the correlations and the dashed and dash-dotted lines indicate the
imaginary parts.  The wavelengths of the oscillations for the light and the heavy FFLO phases are consistent with the formula $\lambda = 2\pi/|q_{\uparrow}+q_{\downarrow}|$. The correlators have been normalized by $n^2$, where $n=n_{\uparrow}+n_{\downarrow}$.}
\end{figure}

Figure \ref{fig:cpair} is a plot of the pair correlation function, which is an indicator of superfluidity.
Indeed, the fully-paired state shows a behavior that is typical of a conventional BCS type of superfluid. The nonzero asymptotics is due the unfixed particle number in the cMPS variational ansatz, similar to a grand-canonical ensemble calculation in which the system is considered in contact with a particle reservoir.
On the other hand, for both the light and heavy partially polarized gases, the correlations show persistent spatial oscillations, displaying the characteristics of a Fulde-Ferrell (FF) type of superfluid with the medium-distance finite-range behavior $\sim e^{i(q_{\uparrow}+q_{\downarrow})x}$ (cf.~Ref.~\cite{Yang2001}).  Note the absence of the expected long-distance algebraic decay. It was replaced by an exponential decay due to the MPS nature of the ansatz, which captures well the local properties, but requires a finite-$D$ scaling analysis to recover the correct long-distance behaviors in the infinite-$D$ limit. 
For lattice systems, this is the complementary situation to what is found with other tensor-network states like the multiscale entanglement renormalization ansatz (MERA) (a continuum version of which is being actively pursued \cite{Haegeman2013c}), which are better adapted to the study of critical properties \cite{Evenbly2013}.

\subsection{Phase Separation}\label{section:PS}
\begin{figure}
\includegraphics[width=\columnwidth]{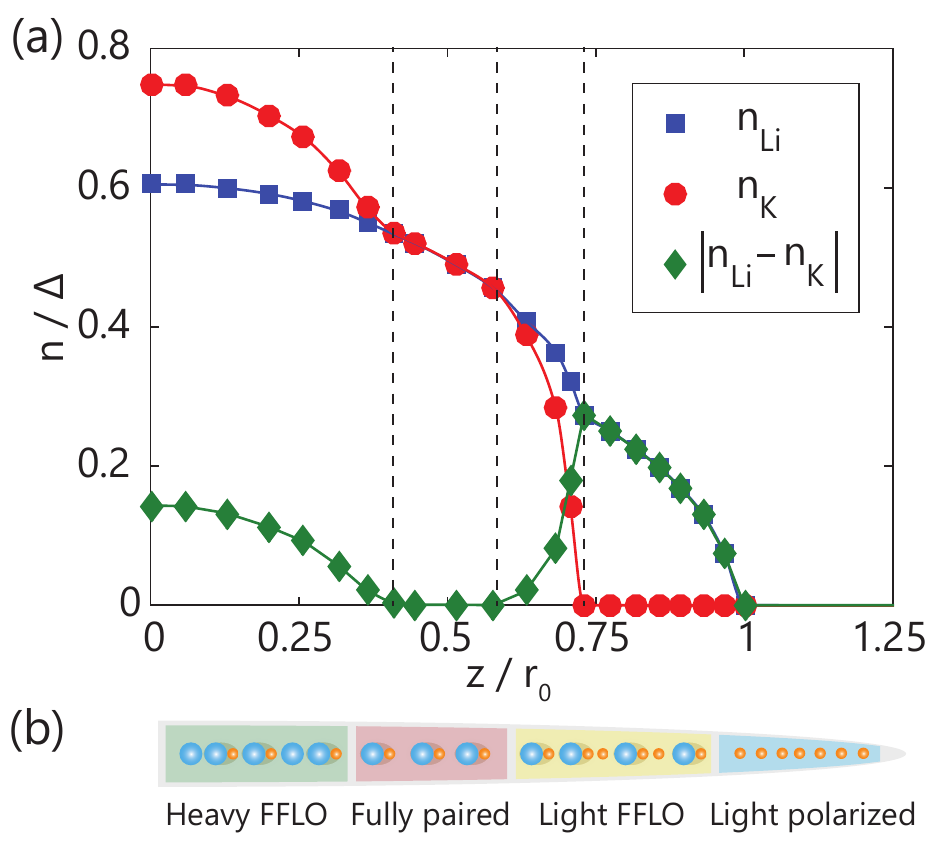}
\caption{\label{fig:tube} (a) Axial density profile of an attractive, spin-imbalanced \LiK gas in a 1D tube of semi-length $r_0$ for an effective magnetic field value of $h=1.9\Delta^2$ and an axial harmonic potential $U(z)=3\Delta^2(z/r_0)^2$, where $\Delta$ is the interaction strength.  The dashed lines indicate the boundaries between different phases occurring.  The densities have been obtained using a local-density approximation, where the chemical potential values range from $\mu = \Delta^2$ (center of tube) to $-2\Delta^2$ (edge of tube).  (b) Pictorial representation of the four distinct phases simultaneously coexisting inside a 1D tube.}
\end{figure}

Figure \ref{fig:tube}(a) is a theoretical density profile of an attractive \LiK gas in a 1D tube, which is essentially a 1D gas confined by an axial harmonic potential.
Within a local-density approximation, it corresponds to a vertical cut on the right side of Fig.~\ref{fig:pd} (going through the point marked as Y). Experimentally, this configuration can be realized, for instance, with 1D tubes of ultracold atoms arranged in a 2D optical lattice as was done in Refs.~\cite{Moritz2005,Liao2010}.  Our prediction for the imbalanced mixture of \LiK gas atoms shows a four-shell structure in 1D tubes.  As depicted graphically in Fig.~\ref{fig:tube}(b), there are three phase separations across the semilength of the 1D tube, where its core is in the heavy-FFLO state, followed sequentially by the fully paired, the light-FFLO and the light-polarized gas shells.  Contrary to 
the case of mass-imbalanced systems in one dimension, only one outer shell of either fully paired or fully polarized gas has been observed outside a partially polarized core for the two-spin \LiSix mixture in one dimension \cite{Liao2010}.  In three dimensions, at most three concentric shell structures have been observed experimentally \cite{Zwierlein2006,Partridge2006,Partridge2006b,Baksmaty2011}, a fully paired superfluid core, followed (in special circumstances) by a partially polarized shell, and surrounded by an outermost shell of a polarized gas, while certain theoretical works predict the possibility of the coexistence of more than three phases \cite{Son2006,Bulgac2006}.

Let us provide a simple physical picture for the emergence of a four-shell structure starting by considering a 1D system of two-component, mass-imbalanced, noninteracting Fermi gas confined by a harmonic potential.  The probability density of finding each particle exponentially decays as it goes away from the center of the trap and the decay length is greater for the lighter species.  Therefore, in a system with a low global polarization, one could expect the density of the heavier particles to be higher than the lighter ones near the center and vice versa away from the center.  The main noticeable effects, densitywise, of an attractive interaction would be the emergence of a finite fully paired region that would have been a single point along the $z$ axis for the noninteracting system and the appearance of fully polarized wings. (Of course superfluidity is another consequence, as discussed below, but that requires the measurement of correlations.)

Figure \ref{fig:pd} indicates that the four-shell structure can be found with an effective field $h>\Delta^2$ and a harmonic potential having a sufficient depth.  In specifications similar to those in past mass-balanced ultracold-atoms experiments, the density profile in Fig.~\ref{fig:tube}(a), for instance, is predicted to be realizable with a total number of approximately 250 atoms in a 1D tube of length 100 $\mu$m, polarization $P=(N_{\uparrow}-N_{\downarrow})/(N_{\uparrow}+N_{\downarrow})$ of $2.5\%$, and interaction strength $\Delta\approx4.2\times 10^{-9}$.

\section{Method}

\subsection{Continuous Matrix Product States}

Our results have been obtained using the cMPS variational ansatz for a system of two-component fermions in a 1D ring or segment of length $L$, defined as
\begin{equation} \left|\Psi\right\rangle
=\mathrm{Tr}_{\mathrm{aux}}\left[\mathcal{P}e^{\int_{0}^{L}dx\left[\mathcal{Q}(x)\otimes\hat{I}+\sum_{\sigma}\mathcal{R}_{\sigma}(x)\otimes\psi_{\sigma}^{\dagger}(x)\right]}\right]\left|\Omega\right\rangle,
\end{equation} 
 where $\Qcal(x)$, $\Rcal{_\sigma}(x)\in\mathbb{C}^{D\times D}$ and act on a $D$-dimensional auxiliary space, $\hat{I}$ is the identity operator on the Fock-space, $\mathrm{Tr_{aux}}$ is a trace over the auxiliary space, $\mathcal{P}$exp is a path-ordered exponential, and $\left|\Omega\right\rangle$ is the Fock-space vacuum state. We follow the approach that we have reported in Ref.~\cite{Chung2015} and we will outline just the modifications applied to our original ansatz.  We will work on the large-$L$ limit, in which the form of the cMPS ansatz can be independent of the details of the boundary conditions on the system \cite{Mei2016}.  In order to accommodate the possibility of spontaneous symmetry breaking properties such as charge density waves (CDWs) and/or spin density waves (SDWs), we introduce new variational parameters $\alpha_{\sigma}$ and $\beta_{\sigma}$ and replace the $\mathcal{R}_\sigma$ matrices to have the spatial dependence
\begin{equation}
\mathcal{R}_{\sigma}(x)=R_{\sigma}\left(\cos\alpha_{\sigma}e^{iq_{\sigma}x}+\sin\alpha_{\sigma}e^{-i\left(q_{\sigma}x+\beta_{\sigma}\right)}\right)\,,\label{eq:Rmatrices}
\end{equation}
 extending our originally proposed plane-wave ansatz $\mathcal{R}_{\sigma}(x)=R_{\sigma}e^{iq_{\sigma}x}$.  The spatially constant matrices $R_{\sigma}$ should satisfy the regularity conditions $\left\{ R_{\uparrow},R_{\downarrow}\right\} =0$ and $R_{\sigma}^{2}=0$. When $\alpha_{\sigma}=n\pi/2$ ($n\in \mathbb{Z}$), we recover the phase modulated $\mathcal{R}_{\sigma}(x)$. This expression for $\mathcal{R}_{\sigma}(x)$ gives the density of species as, 
\begin{equation}
\left\langle \hat{n}_{\sigma}(x)\right\rangle =
\overline{n_{\sigma}}\left[1+\sin2\alpha_{\sigma}\cos\left(2q_{\sigma}x+
\beta_{\sigma}\right)\right]
\end{equation}
 where $\overline{n_{\sigma}}$ is the spatial average of $\langle\hat{n}_{\sigma}(x)\rangle$.  Using the gauge freedom \cite{Haegeman2013a} of the cMPS, we choose $\mathcal{Q}(x)=iH-\frac12\sum_{\sigma}\mathcal{R}^\dagger_{\sigma}(x)\mathcal{R}_{\sigma}(x)$, where the variational Hermitian matrix $H$ is chosen to be spatially independent for simplicity.  The expressions for the expectation values are
\begin{eqnarray}
\langle \psi_{\sigma}^{\dagger}(x)\psi_{\sigma}(x)\rangle &=&  \mathrm{Tr}[\mathcal{P}e^{\int_{0}^{x}dyT(y)}r_{\sigma}(x)\mathcal{P}e^{\int_{x}^{L}dyT(y)}],\nonumber\\
\langle \partial_{x}\hat{\psi}_{\sigma}^{\dagger}(x)\partial_{x}\hat{\psi_{\sigma}}(x)\rangle  &=& \mathrm{Tr}[\mathcal{P}e^{\int_{0}^{x}dyT(y)}t_{\sigma}(x)\mathcal{P}e^{\int_{x}^{L}dyT(y)}],\\\label{eq:expectation}
C_{\mathrm{pair}}(x) &=& \mathrm{Tr}[c_{1}(0)\mathcal{P}e^{\int_{0}^{x}dyT(y)}c_{2}(x)\mathcal{P}e^{\int_{x}^{L}dyT(y)}],\nonumber
\end{eqnarray}
where
\begin{eqnarray}
r_{\sigma}(x)&\equiv&\mathcal{R}_{\sigma}(x)\otimes\bar{\mathcal{R}}_{\sigma}(x),\nonumber\\
T(x)&\equiv&\mathcal{Q}(x)\otimes I+I\otimes\bar{\mathcal{Q}}(x)+\sum_{\sigma}r_{\sigma}(x),\nonumber\\ t_{\sigma}(x)&\equiv&\left\{\partial_{x}\mathcal{R}_{\sigma}(x)+\left[\mathcal{Q}(x),\mathcal{R}_{\sigma}(x)\right]\right\}\otimes\mathrm{c.c.},\\
c_{1}(x) & \equiv & I\otimes\bar{\mathcal{R}}_{\uparrow}(x)\bar{\mathcal{R}}_{\downarrow}(x),\nonumber\\
c_{2}(x) & \equiv & \mathcal{R}_{\uparrow}(x)\mathcal{R}_{\downarrow}(x)\otimes I,\nonumber
\end{eqnarray}
and $I$ is a $D\times D$ identity matrix.

Contrary to our earlier work \cite{Chung2015}, our ansatz breaks the translational invariance of the matrix $T(x)$. This results in expressions that contain exponentials of integrals of $T(x)$ as in Eq.~(\ref{eq:expectation}), instead of simplified expressions, such as $e^{T(L-x)}$. Although this generalization is intuitive, we will show a simple derivation of the expression for the norm of $\left|\chi\right\rangle $ using the standard MPS formalism. The
fermionic cMPS can be written in a discretized form \cite{Verstraete2010} with $N=L/\epsilon$,
\begin{eqnarray}
\left|\chi\right\rangle  & = & \sum_{i_{1},\ldots,i_{N}=0}^{3}\mathrm{Tr}\left(A_{1}^{i_{1}}\cdots A_{N}^{i_{N}}\right)\left|i_{1}\right\rangle \otimes\cdots\otimes\left|i_{N}\right\rangle , \label{eq:chi}
\end{eqnarray}
where $\epsilon=x_{i+1}-x_{i}$.  Unlike bosonic systems where the number of particles at any given site is unlimited, the dimension of the local Hilbert space for the two-component fermions is 4 (due to the Pauli exclusion principle), 
\begin{eqnarray}
\left|0\right\rangle  & = & \left|\:\right\rangle, \nonumber\\
\left|1\right\rangle  & = & \sqrt{\epsilon}\hat{\psi}_{\uparrow}^{\dagger}\left|\:\right\rangle, \nonumber\\
\left|2\right\rangle  & = & \sqrt{\epsilon}\hat{\psi}_{\downarrow}^{\dagger}\left|\:\right\rangle, \\
\left|3\right\rangle  & = & \epsilon\hat{\psi}_{\uparrow}^{\dagger}\hat{\psi}_{\downarrow}^{\dagger}\left|\:\right\rangle ,\nonumber
\end{eqnarray}
where $\left|\:\right\rangle $ is the unoccupied state, and the matrices
$A$ are related to the cMPS matrices $Q$ and $R_{\sigma}$ as
\begin{eqnarray}
A_{j}^{0} & = & 1+\epsilon Q_{j},\nonumber\\
A_{j}^{1} & = & \sqrt{\epsilon}R_{j\uparrow},\nonumber\\
A_{j}^{2} & = & \sqrt{\epsilon}R_{j\downarrow},\\
A_{j}^{3} & = & \frac{\epsilon}{2}\left(R_{j\uparrow}R_{j\downarrow}-R_{j\downarrow}R_{j\uparrow}\right).\nonumber
\end{eqnarray}
Using the adjoint of (\ref{eq:chi}),
\begin{eqnarray}
\left\langle \chi\right| & = & \sum_{i_{1},\ldots,i_{N}}\left\langle i_{1}\right|\otimes\cdots\otimes\left\langle i_{N}\right|\mathrm{Tr}\left(\bar{A}_{1}^{i_{1}}\cdots\bar{A}_{N}^{i_{N}}\right),
\end{eqnarray}
 the norm of (\ref{eq:chi}) can be computed as
\begin{eqnarray}
\left\langle \chi|\chi\right\rangle  & = & \sum_{i_{1},\ldots,i_{N}}\sum_{j_{1},\ldots,j_{N}}\mathrm{Tr}\left(A_{1}^{i_{1}}\cdots A_{N}^{i_{N}}\right)\mathrm{Tr}\left(\bar{A}_{1}^{j_{1}}\cdots\bar{A}_{N}^{jN}\right)\nonumber \\
 &  & \qquad\times\left(\left\langle j_{1}\right|\otimes\cdots\otimes\left\langle j_{N}\right|\right)\left(\left|i_{1}\right\rangle \otimes\cdots\otimes\left|i_{N}\right\rangle \right)\nonumber \\
 & = & \sum_{i_{1},\ldots,i_{N}}\mathrm{Tr}\left(A_{1}^{i_{1}}\cdots A_{N}^{i_{N}}\right)\mathrm{Tr}\left(\bar{A}_{1}^{i_{1}}\cdots\bar{A}_{N}^{i_{N}}\right)\nonumber \\
 & = & \sum_{i_{1}}\mathrm{Tr}\left(A_{1}^{i_{1}}\otimes\bar{A}_{1}^{i_{1}}\right)\sum_{i_{2}}\mathrm{Tr}\left(A_{2}^{i_{2}}\otimes\bar{A}_{2}^{i_{2}}\right)\cdots\sum_{i_{N}}\mathrm{Tr}\left(A_{N}^{i_{N}}\otimes\bar{A}_{N}^{i_{N}}\right)\nonumber \\
 & = & \exp\left(\epsilon T_{1}\right)\exp\left(\epsilon T_{2}\right)\cdots\exp\left(\epsilon T_{N}\right)+O(\epsilon^{2}),\label{eq:expTL}
\end{eqnarray}
where $T_{i}\equiv I\otimes \bar{Q}_{i}+Q_{i}\otimes I+R_{i\uparrow}\otimes \bar{R}_{i\uparrow}+R_{i\downarrow}\otimes \bar{R}_{i\downarrow}$.
In the limit of $\epsilon\rightarrow0$, the expression (\ref{eq:expTL})
can be written as
\begin{eqnarray}
\exp\left(\epsilon T_{1}\right)\exp\left(\epsilon T_{2}\right)...\exp\left(\epsilon T_{N}\right) & = & \exp\left(\epsilon\sum_{i}T_{i}\right)\\
 & = & \exp\left(\int_{0}^{L}dxT(x)\right),\nonumber
\end{eqnarray}
where the integral $\int_{0}^{L}dxT(x)$ in the exponential reduces to $TL$ in the case where $T(x)=T$.  Our modified ansatz (\ref{eq:Rmatrices}) gives a $T(x)$ that is inhomogeneous in space, but the spatially varying parts of $T(x)$ have simple exponential forms that can be analytically integrated with ease.  Moreover, those spatially varying terms give oscillatory non-extensive contributions to the integral over the system length $L$, which can be discarded in the large-system-size limit.\footnote{In a finite system, the variational parameters $q_{\sigma}$ must be properly quantized to ensure the single valuedness of the cMPS.  With this constraint, the contribution of the oscillating terms to the integral over the system length $L$ also becomes zero.}  In addition, it turns out that those terms do not even appear in the converged solutions for homogeneous systems and are present only when we induce them by introducing an oscillatory field or chemical potential.

\subsection{Computations and Numerics}
In our energy-minimization routine, we have assumed that there is no special relation between $q_{\uparrow}$ and $q_{\downarrow}$.  Such relations may occur, for instance, when $n_{\sigma}$ is an integral multiple of $n_{\bar{\sigma}}$.   While our ansatz is able to describe stable states having CDWs and/or SDWs, we did not find any ground state having CDWs or a SDWs from minimizing Eq.~(\ref{eq:freeenergy}) in the absence of any symmetry-breaking perturbation, different from the reports that used lattice models \cite{Cazalilla2005,Wang2009} that assume much denser systems.\footnote{With a chemical potential that oscillates in space, such as $\mu(x)=\mu_0+\mu_1 \cos(kx)$, we were able to find stable ground states showing CDWs or SDWs.}

\begin{figure} 
\includegraphics[width=\columnwidth]{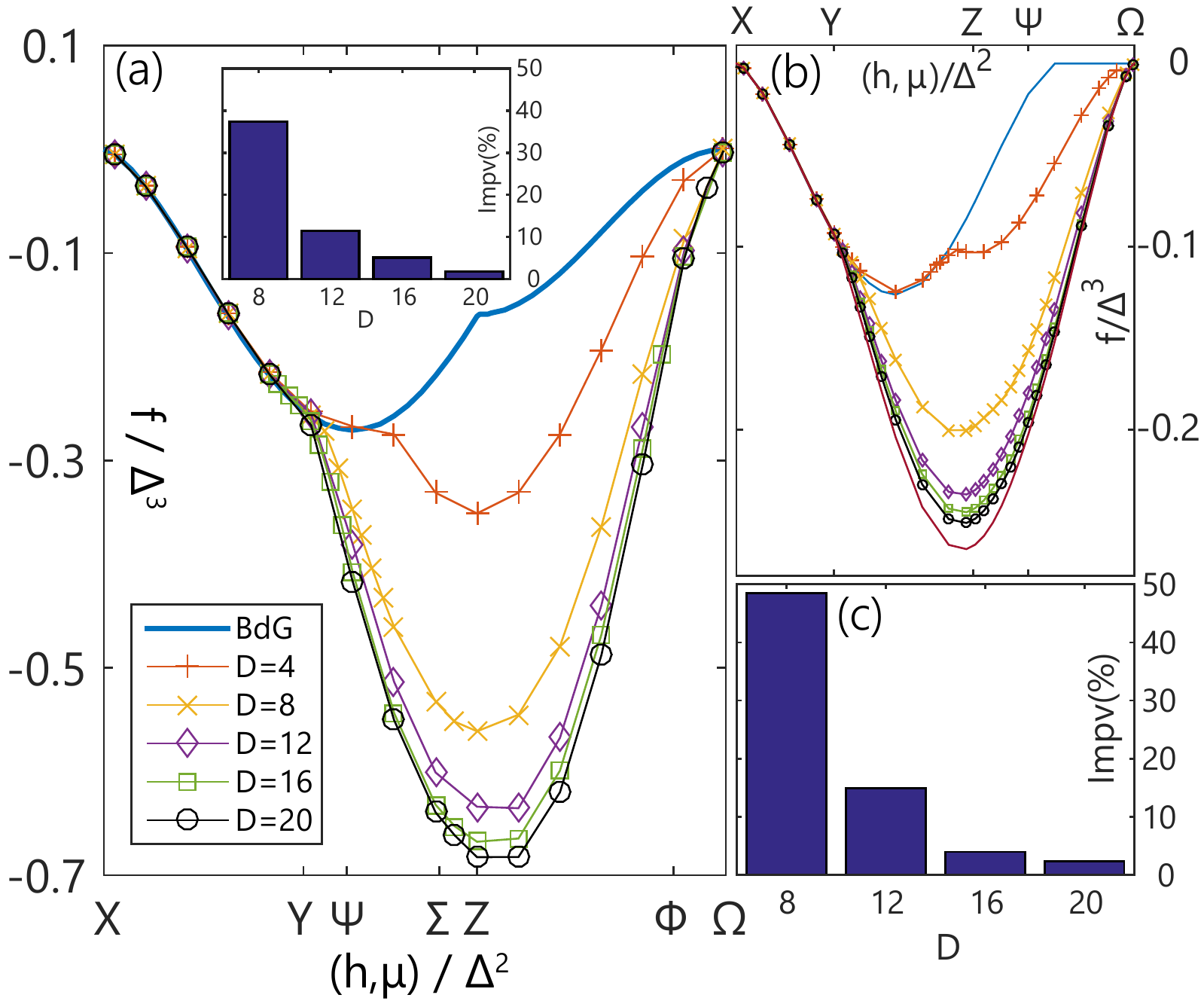}
\caption{\label{fig:energy} (a) Spatial average of the $T=0$ free energy density at the points along circle $\mathcal{C}$ in Fig.~\ref{fig:pd}, obtained from a Bogoliubov--de Gennes
mean-field calculation and the cMPS with $D=4-20$. The inset shows relative improvement of the free-energy-density variational estimate as $D$ is increased, at the point Z in the dashed circle in Fig.~\ref{fig:pd}. The improvement is defined as $|(f_D-f_{D-4})/f_D|$, where $f_D$ is the free energy density at point Z for the bond dimension $D$. (b) Ground-state free-energy density for the mass-balanced system, at the points along the dashed circle $\mathcal{C}$ in the Fig.~\ref{fig:pd} inset. The solid brown curve is the exact result obtained from the Bethe ansatz. (c) Relative improvement of the ground-state free-energy density for the mass-balanced system, at the point Z on circle $\mathcal{C}$ in the inset of Fig.~\ref{fig:pd}.}
\end{figure}

Figure \ref{fig:energy}(a) is a plot of the spatial average of the ground-state free-energy density along the circle $\mathcal{C}$ in the phase diagram (Fig.~\ref{fig:pd}) from cMPSs with bond dimensions ranging from 4 to 20; Bogoliubov-de Gennes (BdG) mean-field calculations are also shown for comparison.
At the fully polarized regime (from point X to Y), which is equivalent to a noninteracting single-species Fermi gas, the BdG free-energy approximations and all bond dimensions of cMPSs agree and coincide.  On the other hand, there are notable variational improvements in the free energy at other regimes in going from BdG to cMPS calculations and while increasing bond dimension. Moreover, it is evident that the cMPS variational estimates are converging and the improvements that result from increasing $D$ are already very small for $D\ge 20$, as can be seen in the inset of Fig.~\ref{fig:energy}(a).

We have also plotted the free-energy density for the mass-balanced system in Fig.~\ref{fig:energy}(b) along the circle $\mathcal{C}$ in the inset of Fig.~\ref{fig:pd}.  An exact solution is available for the equal-mass case, drawn as a solid brown curve, and provides a validation test of convergence for the cMPS results. The corresponding energy-improvement plot at point Z is shown in Fig.~\ref{fig:energy}(c).  We see that the trend of energy improvement (as a function of $D$) for the mass-imbalanced system is similar to that of the mass-balanced system.  From this, we can infer the reliability of our findings (the phase diagram and the prediction of the four-shell structure) for the \LiK system to be comparable to that of our results for the mass-balanced system, which we have discussed in detail in Ref. \cite{Chung2015}.  The computational cost for each data point (with $D=16$) was on the order of a day (wall time), on a single Intel Xeon X5650 2.66 GHz processor with an Nvidia Tesla M2070 graphical processing unit (GPU).  Notice that fermionic cMPSs are notoriously more difficult to optimize than bosonic ones.  A significantly improved optimization algorithm for the bosonic case has been reported recently \cite{Ganahl2017} and it would be an interesting question whether those ideas can be efficiently applied also to our two-component fermionic ansatz.

\section{Conclusions and Outlook}

We have used a fermionic cMPS variational ansatz to compute the ground state phases of the attractive \LiK gas in one dimension as a function of the effective chemical potential and the effective magnetic field.  We have argued the validity of our calculations by comparing the free-energy densities from various bond dimensions (and also the BdG mean-field results).  The two partially polarized states (light and heavy) were found to be inhomogeneous superfluids of the FFLO type.   Using a local-density approximation, we predict that the \LiK system, realized, for instance, with a 2D optical lattice superimposed on a harmonic trap, could exhibit an intriguing four-shell structure along 1D tubes of \LiK gas, simultaneously displaying light-polarized, light-FFLO, fully paired, and heavy-FFLO phases, thus displaying higher complexity than the mass-balanced case where only two-shell structures are predicted and seen.  Finding oppositely polarized phases at the center and the edges of the 1D tubes would be the immediate initial hallmark for experimental tests of our predictions.  This application of the cMPS on a nonintegrable model demonstrates its potential as a versatile numerical tool for systems of ultracold atomic gases confined to one dimension.

\begin{acknowledgments}
We acknowledge discussions with J. I. Cirac, T. Giarmarchi, J. Haegeman, K. Sun, and F. Verstraete.  We are grateful for the hospitality of the Kavli Institute for Theoretical Physics at UCSB (funded under Grant No. NSF PHY-1125915) and IISER-Pune, where some of the writing took place. The auspices of the NSF (Grant No. PHY-1708049) via the AMOT and CMMT programs are acknowledged. Finally, we are grateful for the support by the Ohio Supercomputer Center \cite{OhioSupercomputerCenter1987} where the computations were done.
\end{acknowledgments}


\begin{thebibliography}{71}%
\makeatletter
\providecommand \@ifxundefined [1]{%
 \@ifx{#1\undefined}
}%
\providecommand \@ifnum [1]{%
 \ifnum #1\expandafter \@firstoftwo
 \else \expandafter \@secondoftwo
 \fi
}%
\providecommand \@ifx [1]{%
 \ifx #1\expandafter \@firstoftwo
 \else \expandafter \@secondoftwo
 \fi
}%
\providecommand \natexlab [1]{#1}%
\providecommand \enquote  [1]{``#1''}%
\providecommand \bibnamefont  [1]{#1}%
\providecommand \bibfnamefont [1]{#1}%
\providecommand \citenamefont [1]{#1}%
\providecommand \href@noop [0]{\@secondoftwo}%
\providecommand \href [0]{\begingroup \@sanitize@url \@href}%
\providecommand \@href[1]{\@@startlink{#1}\@@href}%
\providecommand \@@href[1]{\endgroup#1\@@endlink}%
\providecommand \@sanitize@url [0]{\catcode `\\12\catcode `\$12\catcode
  `\&12\catcode `\#12\catcode `\^12\catcode `\_12\catcode `\%12\relax}%
\providecommand \@@startlink[1]{}%
\providecommand \@@endlink[0]{}%
\providecommand \url  [0]{\begingroup\@sanitize@url \@url }%
\providecommand \@url [1]{\endgroup\@href {#1}{\urlprefix }}%
\providecommand \urlprefix  [0]{URL }%
\providecommand \Eprint [0]{\href }%
\providecommand \doibase [0]{http://dx.doi.org/}%
\providecommand \selectlanguage [0]{\@gobble}%
\providecommand \bibinfo  [0]{\@secondoftwo}%
\providecommand \bibfield  [0]{\@secondoftwo}%
\providecommand \translation [1]{[#1]}%
\providecommand \BibitemOpen [0]{}%
\providecommand \bibitemStop [0]{}%
\providecommand \bibitemNoStop [0]{.\EOS\space}%
\providecommand \EOS [0]{\spacefactor3000\relax}%
\providecommand \BibitemShut  [1]{\csname bibitem#1\endcsname}%
\let\auto@bib@innerbib\@empty
\bibitem [{\citenamefont {Giorgini}\ \emph {et~al.}(2008)\citenamefont
  {Giorgini}, \citenamefont {Pitaevskii},\ and\ \citenamefont
  {Stringari}}]{Giorgini2008}%
  \BibitemOpen
  \bibfield  {author} {\bibinfo {author} {\bibfnamefont {S.}~\bibnamefont
  {Giorgini}}, \bibinfo {author} {\bibfnamefont {L.}~\bibnamefont
  {Pitaevskii}}, \ and\ \bibinfo {author} {\bibfnamefont {S.}~\bibnamefont
  {Stringari}},\ }\bibfield  {title} {\enquote {\bibinfo {title} {Theory of
  ultracold atomic {F}ermi gases},}\ }\href {\doibase
  10.1103/RevModPhys.80.1215} {\bibfield  {journal} {\bibinfo  {journal} {Rev.
  Mod. Phys.}\ }\textbf {\bibinfo {volume} {80}},\ \bibinfo {pages}
  {1215--1274} (\bibinfo {year} {2008})}\BibitemShut {NoStop}%
\bibitem [{\citenamefont {Guan}\ \emph {et~al.}(2013)\citenamefont {Guan},
  \citenamefont {Batchelor},\ and\ \citenamefont {Lee}}]{Guan2013}%
  \BibitemOpen
  \bibfield  {author} {\bibinfo {author} {\bibfnamefont {X.-W.}\ \bibnamefont
  {Guan}}, \bibinfo {author} {\bibfnamefont {M.~T.}\ \bibnamefont {Batchelor}},
  \ and\ \bibinfo {author} {\bibfnamefont {C.}~\bibnamefont {Lee}},\ }\bibfield
   {title} {\enquote {\bibinfo {title} {{F}ermi gases in one dimension: From
  {B}ethe ansatz to experiments},}\ }\href {\doibase
  10.1103/RevModPhys.85.1633} {\bibfield  {journal} {\bibinfo  {journal} {Rev.
  Mod. Phys.}\ }\textbf {\bibinfo {volume} {85}},\ \bibinfo {pages}
  {1633--1691} (\bibinfo {year} {2013})}\BibitemShut {NoStop}%
\bibitem [{\citenamefont {Gubbels}\ and\ \citenamefont
  {Stoof}(2013)}]{Gubbels2013}%
  \BibitemOpen
  \bibfield  {author} {\bibinfo {author} {\bibfnamefont {K.~B.}\ \bibnamefont
  {Gubbels}}\ and\ \bibinfo {author} {\bibfnamefont {H.~T.~C.}\ \bibnamefont
  {Stoof}},\ }\bibfield  {title} {\enquote {\bibinfo {title} {Imbalanced
  {F}ermi gases at unitarity},}\ }\href {\doibase
  10.1016/j.physrep.2012.11.004} {\bibfield  {journal} {\bibinfo  {journal}
  {Physics Reports}\ }\textbf {\bibinfo {volume} {525}},\ \bibinfo {pages} {255
  -- 313} (\bibinfo {year} {2013})}\BibitemShut {NoStop}%
\bibitem [{\citenamefont {Zwierlein}\ \emph {et~al.}(2004)\citenamefont
  {Zwierlein}, \citenamefont {Stan}, \citenamefont {Schunck}, \citenamefont
  {Raupach}, \citenamefont {Kerman},\ and\ \citenamefont
  {Ketterle}}]{Zwierlein2004}%
  \BibitemOpen
  \bibfield  {author} {\bibinfo {author} {\bibfnamefont {M.~W.}\ \bibnamefont
  {Zwierlein}}, \bibinfo {author} {\bibfnamefont {C.~A.}\ \bibnamefont {Stan}},
  \bibinfo {author} {\bibfnamefont {C.~H.}\ \bibnamefont {Schunck}}, \bibinfo
  {author} {\bibfnamefont {S.~M.~F.}\ \bibnamefont {Raupach}}, \bibinfo
  {author} {\bibfnamefont {A.~J.}\ \bibnamefont {Kerman}}, \ and\ \bibinfo
  {author} {\bibfnamefont {W.}~\bibnamefont {Ketterle}},\ }\bibfield  {title}
  {\enquote {\bibinfo {title} {Condensation of pairs of fermionic atoms near a
  {F}eshbach resonance},}\ }\href {\doibase 10.1103/PhysRevLett.92.120403}
  {\bibfield  {journal} {\bibinfo  {journal} {Phys. Rev. Lett.}\ }\textbf
  {\bibinfo {volume} {92}},\ \bibinfo {pages} {120403} (\bibinfo {year}
  {2004})}\BibitemShut {NoStop}%
\bibitem [{\citenamefont {Regal}\ \emph {et~al.}(2004)\citenamefont {Regal},
  \citenamefont {Greiner},\ and\ \citenamefont {Jin}}]{Regal2004}%
  \BibitemOpen
  \bibfield  {author} {\bibinfo {author} {\bibfnamefont {C.~A.}\ \bibnamefont
  {Regal}}, \bibinfo {author} {\bibfnamefont {M.}~\bibnamefont {Greiner}}, \
  and\ \bibinfo {author} {\bibfnamefont {D.~S.}\ \bibnamefont {Jin}},\
  }\bibfield  {title} {\enquote {\bibinfo {title} {Observation of resonance
  condensation of fermionic atom pairs},}\ }\href {\doibase
  10.1103/PhysRevLett.92.040403} {\bibfield  {journal} {\bibinfo  {journal}
  {Phys. Rev. Lett.}\ }\textbf {\bibinfo {volume} {92}},\ \bibinfo {pages}
  {040403} (\bibinfo {year} {2004})}\BibitemShut {NoStop}%
\bibitem [{\citenamefont {Bartenstein}\ \emph {et~al.}(2004)\citenamefont
  {Bartenstein}, \citenamefont {Altmeyer}, \citenamefont {Riedl}, \citenamefont
  {Jochim}, \citenamefont {Chin}, \citenamefont {Denschlag},\ and\
  \citenamefont {Grimm}}]{Bartenstein2004}%
  \BibitemOpen
  \bibfield  {author} {\bibinfo {author} {\bibfnamefont {M.}~\bibnamefont
  {Bartenstein}}, \bibinfo {author} {\bibfnamefont {A.}~\bibnamefont
  {Altmeyer}}, \bibinfo {author} {\bibfnamefont {S.}~\bibnamefont {Riedl}},
  \bibinfo {author} {\bibfnamefont {S.}~\bibnamefont {Jochim}}, \bibinfo
  {author} {\bibfnamefont {C.}~\bibnamefont {Chin}}, \bibinfo {author}
  {\bibfnamefont {J.~H.}\ \bibnamefont {Denschlag}}, \ and\ \bibinfo {author}
  {\bibfnamefont {R.}~\bibnamefont {Grimm}},\ }\bibfield  {title} {\enquote
  {\bibinfo {title} {Collective excitations of a degenerate gas at the
  {BEC}-{BCS} crossover},}\ }\href {\doibase 10.1103/PhysRevLett.92.203201}
  {\bibfield  {journal} {\bibinfo  {journal} {Phys. Rev. Lett.}\ }\textbf
  {\bibinfo {volume} {92}},\ \bibinfo {pages} {203201} (\bibinfo {year}
  {2004})}\BibitemShut {NoStop}%
\bibitem [{\citenamefont {Kinast}\ \emph {et~al.}(2004)\citenamefont {Kinast},
  \citenamefont {Hemmer}, \citenamefont {Gehm}, \citenamefont {Turlapov},\ and\
  \citenamefont {Thomas}}]{Kinast2004}%
  \BibitemOpen
  \bibfield  {author} {\bibinfo {author} {\bibfnamefont {J.}~\bibnamefont
  {Kinast}}, \bibinfo {author} {\bibfnamefont {S.~L.}\ \bibnamefont {Hemmer}},
  \bibinfo {author} {\bibfnamefont {M.~E.}\ \bibnamefont {Gehm}}, \bibinfo
  {author} {\bibfnamefont {A.}~\bibnamefont {Turlapov}}, \ and\ \bibinfo
  {author} {\bibfnamefont {J.~E.}\ \bibnamefont {Thomas}},\ }\bibfield  {title}
  {\enquote {\bibinfo {title} {Evidence for superfluidity in a resonantly
  interacting {F}ermi gas},}\ }\href {\doibase 10.1103/PhysRevLett.92.150402}
  {\bibfield  {journal} {\bibinfo  {journal} {Phys. Rev. Lett.}\ }\textbf
  {\bibinfo {volume} {92}},\ \bibinfo {pages} {150402} (\bibinfo {year}
  {2004})}\BibitemShut {NoStop}%
\bibitem [{\citenamefont {Partridge}\ \emph {et~al.}(2005)\citenamefont
  {Partridge}, \citenamefont {Strecker}, \citenamefont {Kamar}, \citenamefont
  {Jack},\ and\ \citenamefont {Hulet}}]{Partridge2005}%
  \BibitemOpen
  \bibfield  {author} {\bibinfo {author} {\bibfnamefont {G.~B.}\ \bibnamefont
  {Partridge}}, \bibinfo {author} {\bibfnamefont {K.~E.}\ \bibnamefont
  {Strecker}}, \bibinfo {author} {\bibfnamefont {R.~I.}\ \bibnamefont {Kamar}},
  \bibinfo {author} {\bibfnamefont {M.~W.}\ \bibnamefont {Jack}}, \ and\
  \bibinfo {author} {\bibfnamefont {R.~G.}\ \bibnamefont {Hulet}},\ }\bibfield
  {title} {\enquote {\bibinfo {title} {Molecular probe of pairing in the
  {BEC}-{BCS} crossover},}\ }\href {\doibase 10.1103/PhysRevLett.95.020404}
  {\bibfield  {journal} {\bibinfo  {journal} {Phys. Rev. Lett.}\ }\textbf
  {\bibinfo {volume} {95}},\ \bibinfo {pages} {020404} (\bibinfo {year}
  {2005})}\BibitemShut {NoStop}%
\bibitem [{\citenamefont {Zwierlein}\ \emph {et~al.}(2005)\citenamefont
  {Zwierlein}, \citenamefont {Abo-Shaeer}, \citenamefont {Schirotzek},
  \citenamefont {Schunck},\ and\ \citenamefont {Ketterle}}]{Zwierlein2005}%
  \BibitemOpen
  \bibfield  {author} {\bibinfo {author} {\bibfnamefont {M.~W.}\ \bibnamefont
  {Zwierlein}}, \bibinfo {author} {\bibfnamefont {J.~R.}\ \bibnamefont
  {Abo-Shaeer}}, \bibinfo {author} {\bibfnamefont {A.}~\bibnamefont
  {Schirotzek}}, \bibinfo {author} {\bibfnamefont {C.~H.}\ \bibnamefont
  {Schunck}}, \ and\ \bibinfo {author} {\bibfnamefont {W.}~\bibnamefont
  {Ketterle}},\ }\bibfield  {title} {\enquote {\bibinfo {title} {Vortices and
  superfluidity in a strongly interacting {F}ermi gas},}\ }\href {\doibase
  10.1038/nature03858} {\bibfield  {journal} {\bibinfo  {journal} {Nature}\
  }\textbf {\bibinfo {volume} {435}},\ \bibinfo {pages} {1047--1051} (\bibinfo
  {year} {2005})}\BibitemShut {NoStop}%
\bibitem [{\citenamefont {Zwierlein}\ \emph {et~al.}(2006)\citenamefont
  {Zwierlein}, \citenamefont {Schirotzek}, \citenamefont {Schunck},\ and\
  \citenamefont {Ketterle}}]{Zwierlein2006}%
  \BibitemOpen
  \bibfield  {author} {\bibinfo {author} {\bibfnamefont {M.~W.}\ \bibnamefont
  {Zwierlein}}, \bibinfo {author} {\bibfnamefont {A.}~\bibnamefont
  {Schirotzek}}, \bibinfo {author} {\bibfnamefont {C.~H.}\ \bibnamefont
  {Schunck}}, \ and\ \bibinfo {author} {\bibfnamefont {W.}~\bibnamefont
  {Ketterle}},\ }\bibfield  {title} {\enquote {\bibinfo {title} {Fermionic
  superfluidity with imbalanced spin populations},}\ }\href {\doibase
  10.1126/science.1122318} {\bibfield  {journal} {\bibinfo  {journal}
  {Science}\ }\textbf {\bibinfo {volume} {311}},\ \bibinfo {pages} {492--496}
  (\bibinfo {year} {2006})}\BibitemShut {NoStop}%
\bibitem [{\citenamefont {Partridge}\ \emph
  {et~al.}(2006{\natexlab{a}})\citenamefont {Partridge}, \citenamefont {Li},
  \citenamefont {Kamar}, \citenamefont {Liao},\ and\ \citenamefont
  {Hulet}}]{Partridge2006}%
  \BibitemOpen
  \bibfield  {author} {\bibinfo {author} {\bibfnamefont {G.~B.}\ \bibnamefont
  {Partridge}}, \bibinfo {author} {\bibfnamefont {W.}~\bibnamefont {Li}},
  \bibinfo {author} {\bibfnamefont {R.~I.}\ \bibnamefont {Kamar}}, \bibinfo
  {author} {\bibfnamefont {Y.-a.}\ \bibnamefont {Liao}}, \ and\ \bibinfo
  {author} {\bibfnamefont {R.~G.}\ \bibnamefont {Hulet}},\ }\bibfield  {title}
  {\enquote {\bibinfo {title} {Pairing and phase separation in a polarized
  {F}ermi gas},}\ }\href {\doibase 10.1126/science.1122876} {\bibfield
  {journal} {\bibinfo  {journal} {Science}\ }\textbf {\bibinfo {volume}
  {311}},\ \bibinfo {pages} {503--505} (\bibinfo {year}
  {2006}{\natexlab{a}})}\BibitemShut {NoStop}%
\bibitem [{\citenamefont {Partridge}\ \emph
  {et~al.}(2006{\natexlab{b}})\citenamefont {Partridge}, \citenamefont {Li},
  \citenamefont {Liao}, \citenamefont {Hulet}, \citenamefont {Haque},\ and\
  \citenamefont {Stoof}}]{Partridge2006b}%
  \BibitemOpen
  \bibfield  {author} {\bibinfo {author} {\bibfnamefont {G.~B.}\ \bibnamefont
  {Partridge}}, \bibinfo {author} {\bibfnamefont {W.}~\bibnamefont {Li}},
  \bibinfo {author} {\bibfnamefont {Y.~A.}\ \bibnamefont {Liao}}, \bibinfo
  {author} {\bibfnamefont {R.~G.}\ \bibnamefont {Hulet}}, \bibinfo {author}
  {\bibfnamefont {M.}~\bibnamefont {Haque}}, \ and\ \bibinfo {author}
  {\bibfnamefont {H.~T.~C.}\ \bibnamefont {Stoof}},\ }\bibfield  {title}
  {\enquote {\bibinfo {title} {Deformation of a trapped {F}ermi gas with
  unequal spin populations},}\ }\href {\doibase 10.1103/PhysRevLett.97.190407}
  {\bibfield  {journal} {\bibinfo  {journal} {Phys. Rev. Lett.}\ }\textbf
  {\bibinfo {volume} {97}},\ \bibinfo {pages} {190407} (\bibinfo {year}
  {2006}{\natexlab{b}})}\BibitemShut {NoStop}%
\bibitem [{\citenamefont {Liao}\ \emph {et~al.}(2010)\citenamefont {Liao},
  \citenamefont {Rittner}, \citenamefont {Paprotta}, \citenamefont {Li},
  \citenamefont {Partridge}, \citenamefont {Hulet}, \citenamefont {Baur},\ and\
  \citenamefont {Mueller}}]{Liao2010}%
  \BibitemOpen
  \bibfield  {author} {\bibinfo {author} {\bibfnamefont {Y.-A.}\ \bibnamefont
  {Liao}}, \bibinfo {author} {\bibfnamefont {A.~S.~C.}\ \bibnamefont
  {Rittner}}, \bibinfo {author} {\bibfnamefont {T.}~\bibnamefont {Paprotta}},
  \bibinfo {author} {\bibfnamefont {W.}~\bibnamefont {Li}}, \bibinfo {author}
  {\bibfnamefont {G.~B.}\ \bibnamefont {Partridge}}, \bibinfo {author}
  {\bibfnamefont {R.~G}\ \bibnamefont {Hulet}}, \bibinfo {author}
  {\bibfnamefont {S.~K.}\ \bibnamefont {Baur}}, \ and\ \bibinfo {author}
  {\bibfnamefont {E.~J.}\ \bibnamefont {Mueller}},\ }\bibfield  {title}
  {\enquote {\bibinfo {title} {Spin-imbalance in a one-dimensional {F}ermi
  gas},}\ }\href {\doibase 10.1038/nature09393} {\bibfield  {journal} {\bibinfo
   {journal} {Nature}\ }\textbf {\bibinfo {volume} {467}},\ \bibinfo {pages}
  {567--569} (\bibinfo {year} {2010})}\BibitemShut {NoStop}%
\bibitem [{\citenamefont {Orso}(2007)}]{Orso2007}%
  \BibitemOpen
  \bibfield  {author} {\bibinfo {author} {\bibfnamefont {G.}~\bibnamefont
  {Orso}},\ }\bibfield  {title} {\enquote {\bibinfo {title} {Attractive {F}ermi
  gases with unequal spin populations in highly elongated traps},}\ }\href
  {\doibase 10.1103/PhysRevLett.98.070402} {\bibfield  {journal} {\bibinfo
  {journal} {Phys. Rev. Lett.}\ }\textbf {\bibinfo {volume} {98}},\ \bibinfo
  {pages} {070402} (\bibinfo {year} {2007})}\BibitemShut {NoStop}%
\bibitem [{\citenamefont {Hu}\ \emph {et~al.}(2007)\citenamefont {Hu},
  \citenamefont {Liu},\ and\ \citenamefont {Drummond}}]{Hu2007}%
  \BibitemOpen
  \bibfield  {author} {\bibinfo {author} {\bibfnamefont {H.}~\bibnamefont
  {Hu}}, \bibinfo {author} {\bibfnamefont {X.-J.}\ \bibnamefont {Liu}}, \ and\
  \bibinfo {author} {\bibfnamefont {P.~D.}\ \bibnamefont {Drummond}},\
  }\bibfield  {title} {\enquote {\bibinfo {title} {Phase diagram of a strongly
  interacting polarized {F}ermi gas in one dimension},}\ }\href {\doibase
  10.1103/PhysRevLett.98.070403} {\bibfield  {journal} {\bibinfo  {journal}
  {Phys. Rev. Lett.}\ }\textbf {\bibinfo {volume} {98}},\ \bibinfo {pages}
  {070403} (\bibinfo {year} {2007})}\BibitemShut {NoStop}%
\bibitem [{\citenamefont {Fulde}\ and\ \citenamefont
  {Ferrell}(1964)}]{Fulde1964}%
  \BibitemOpen
  \bibfield  {author} {\bibinfo {author} {\bibfnamefont {P.}~\bibnamefont
  {Fulde}}\ and\ \bibinfo {author} {\bibfnamefont {R.~A.}\ \bibnamefont
  {Ferrell}},\ }\bibfield  {title} {\enquote {\bibinfo {title}
  {Superconductivity in a strong spin-exchange field},}\ }\href {\doibase
  10.1103/PhysRev.135.A550} {\bibfield  {journal} {\bibinfo  {journal} {Phys.
  Rev.}\ }\textbf {\bibinfo {volume} {135}},\ \bibinfo {pages} {A550--A563}
  (\bibinfo {year} {1964})}\BibitemShut {NoStop}%
\bibitem [{\citenamefont {Larkin}\ and\ \citenamefont
  {Ovchinnikov}(1965)}]{Larkin1965}%
  \BibitemOpen
  \bibfield  {author} {\bibinfo {author} {\bibfnamefont {A.~I.}\ \bibnamefont
  {Larkin}}\ and\ \bibinfo {author} {\bibfnamefont {Y.~N.}\ \bibnamefont
  {Ovchinnikov}},\ }\bibfield  {title} {\enquote {\bibinfo {title} {Nonuniform
  state of superconductors},}\ }\href@noop {} {\bibfield  {journal} {\bibinfo
  {journal} {Sov. Phys. JETP}\ }\textbf {\bibinfo {volume} {20}},\ \bibinfo
  {pages} {762} (\bibinfo {year} {1965})}\BibitemShut {NoStop}%
\bibitem [{\citenamefont {Casalbuoni}\ and\ \citenamefont
  {Nardulli}(2004)}]{Casalbuoni2004}%
  \BibitemOpen
  \bibfield  {author} {\bibinfo {author} {\bibfnamefont {R.}~\bibnamefont
  {Casalbuoni}}\ and\ \bibinfo {author} {\bibfnamefont {G.}~\bibnamefont
  {Nardulli}},\ }\bibfield  {title} {\enquote {\bibinfo {title} {Inhomogeneous
  superconductivity in condensed matter and {QCD}},}\ }\href {\doibase
  10.1103/RevModPhys.76.263} {\bibfield  {journal} {\bibinfo  {journal} {Rev.
  Mod. Phys.}\ }\textbf {\bibinfo {volume} {76}},\ \bibinfo {pages} {263--320}
  (\bibinfo {year} {2004})}\BibitemShut {NoStop}%
\bibitem [{\citenamefont {Feiguin}\ and\ \citenamefont
  {Heidrich-Meisner}(2007)}]{Feiguin2007}%
  \BibitemOpen
  \bibfield  {author} {\bibinfo {author} {\bibfnamefont {A.~E.}\ \bibnamefont
  {Feiguin}}\ and\ \bibinfo {author} {\bibfnamefont {F.}~\bibnamefont
  {Heidrich-Meisner}},\ }\bibfield  {title} {\enquote {\bibinfo {title}
  {Pairing states of a polarized {F}ermi gas trapped in a one-dimensional
  optical lattice},}\ }\href {\doibase 10.1103/PhysRevB.76.220508} {\bibfield
  {journal} {\bibinfo  {journal} {Phys. Rev. B}\ }\textbf {\bibinfo {volume}
  {76}},\ \bibinfo {pages} {220508} (\bibinfo {year} {2007})}\BibitemShut
  {NoStop}%
\bibitem [{\citenamefont {Casula}\ \emph {et~al.}(2008)\citenamefont {Casula},
  \citenamefont {Ceperley},\ and\ \citenamefont {Mueller}}]{Casula2008}%
  \BibitemOpen
  \bibfield  {author} {\bibinfo {author} {\bibfnamefont {M.}~\bibnamefont
  {Casula}}, \bibinfo {author} {\bibfnamefont {D.~M.}\ \bibnamefont
  {Ceperley}}, \ and\ \bibinfo {author} {\bibfnamefont {E.~J.}\ \bibnamefont
  {Mueller}},\ }\bibfield  {title} {\enquote {\bibinfo {title} {Quantum {M}onte
  {C}arlo study of one-dimensional trapped fermions with attractive contact
  interactions},}\ }\href {\doibase 10.1103/PhysRevA.78.033607} {\bibfield
  {journal} {\bibinfo  {journal} {Phys. Rev. A}\ }\textbf {\bibinfo {volume}
  {78}},\ \bibinfo {pages} {033607} (\bibinfo {year} {2008})}\BibitemShut
  {NoStop}%
\bibitem [{\citenamefont {Kakashvili}\ and\ \citenamefont
  {Bolech}(2009)}]{Kakashvili2009}%
  \BibitemOpen
  \bibfield  {author} {\bibinfo {author} {\bibfnamefont {P.}~\bibnamefont
  {Kakashvili}}\ and\ \bibinfo {author} {\bibfnamefont {C.~J.}\ \bibnamefont
  {Bolech}},\ }\bibfield  {title} {\enquote {\bibinfo {title} {Paired states in
  spin-imbalanced atomic {F}ermi gases in one dimension},}\ }\href {\doibase
  10.1103/PhysRevA.79.041603} {\bibfield  {journal} {\bibinfo  {journal} {Phys.
  Rev. A}\ }\textbf {\bibinfo {volume} {79}},\ \bibinfo {pages} {041603}
  (\bibinfo {year} {2009})}\BibitemShut {NoStop}%
\bibitem [{\citenamefont {Sun}\ and\ \citenamefont {Bolech}(2013)}]{Sun2013}%
  \BibitemOpen
  \bibfield  {author} {\bibinfo {author} {\bibfnamefont {K.}~\bibnamefont
  {Sun}}\ and\ \bibinfo {author} {\bibfnamefont {C.~J.}\ \bibnamefont
  {Bolech}},\ }\bibfield  {title} {\enquote {\bibinfo {title} {Pair tunneling,
  phase separation, and dimensional crossover in imbalanced fermionic
  superfluids in a coupled array of tubes},}\ }\href {\doibase
  10.1103/PhysRevA.87.053622} {\bibfield  {journal} {\bibinfo  {journal} {Phys.
  Rev. A}\ }\textbf {\bibinfo {volume} {87}},\ \bibinfo {pages} {053622}
  (\bibinfo {year} {2013})}\BibitemShut {NoStop}%
\bibitem [{\citenamefont {Bianchi}\ \emph {et~al.}(2003)\citenamefont
  {Bianchi}, \citenamefont {Movshovich}, \citenamefont {Capan}, \citenamefont
  {Pagliuso},\ and\ \citenamefont {Sarrao}}]{Bianchi2003}%
  \BibitemOpen
  \bibfield  {author} {\bibinfo {author} {\bibfnamefont {A.}~\bibnamefont
  {Bianchi}}, \bibinfo {author} {\bibfnamefont {R.}~\bibnamefont {Movshovich}},
  \bibinfo {author} {\bibfnamefont {C.}~\bibnamefont {Capan}}, \bibinfo
  {author} {\bibfnamefont {P.~G.}\ \bibnamefont {Pagliuso}}, \ and\ \bibinfo
  {author} {\bibfnamefont {J.~L.}\ \bibnamefont {Sarrao}},\ }\bibfield  {title}
  {\enquote {\bibinfo {title} {Possible
  {F}ulde-{F}errell-{L}arkin-{O}vchinnikov superconducting state in
  {CeCoIn}${}_{5}$},}\ }\href {\doibase 10.1103/PhysRevLett.91.187004}
  {\bibfield  {journal} {\bibinfo  {journal} {Phys. Rev. Lett.}\ }\textbf
  {\bibinfo {volume} {91}},\ \bibinfo {pages} {187004} (\bibinfo {year}
  {2003})}\BibitemShut {NoStop}%
\bibitem [{\citenamefont {Uji}\ \emph {et~al.}(2006)\citenamefont {Uji},
  \citenamefont {Terashima}, \citenamefont {Nishimura}, \citenamefont
  {Takahide}, \citenamefont {Konoike}, \citenamefont {Enomoto}, \citenamefont
  {Cui}, \citenamefont {Kobayashi}, \citenamefont {Kobayashi}, \citenamefont
  {Tanaka}, \citenamefont {Tokumoto}, \citenamefont {Choi}, \citenamefont
  {Tokumoto}, \citenamefont {Graf},\ and\ \citenamefont {Brooks}}]{Uji2006}%
  \BibitemOpen
  \bibfield  {author} {\bibinfo {author} {\bibfnamefont {S.}~\bibnamefont
  {Uji}}, \bibinfo {author} {\bibfnamefont {T.}~\bibnamefont {Terashima}},
  \bibinfo {author} {\bibfnamefont {M.}~\bibnamefont {Nishimura}}, \bibinfo
  {author} {\bibfnamefont {Y.}~\bibnamefont {Takahide}}, \bibinfo {author}
  {\bibfnamefont {T.}~\bibnamefont {Konoike}}, \bibinfo {author} {\bibfnamefont
  {K.}~\bibnamefont {Enomoto}}, \bibinfo {author} {\bibfnamefont
  {H.}~\bibnamefont {Cui}}, \bibinfo {author} {\bibfnamefont {H.}~\bibnamefont
  {Kobayashi}}, \bibinfo {author} {\bibfnamefont {A.}~\bibnamefont
  {Kobayashi}}, \bibinfo {author} {\bibfnamefont {H.}~\bibnamefont {Tanaka}},
  \bibinfo {author} {\bibfnamefont {M.}~\bibnamefont {Tokumoto}}, \bibinfo
  {author} {\bibfnamefont {E.~S.}\ \bibnamefont {Choi}}, \bibinfo {author}
  {\bibfnamefont {T.}~\bibnamefont {Tokumoto}}, \bibinfo {author}
  {\bibfnamefont {D.}~\bibnamefont {Graf}}, \ and\ \bibinfo {author}
  {\bibfnamefont {J.~S.}\ \bibnamefont {Brooks}},\ }\bibfield  {title}
  {\enquote {\bibinfo {title} {Vortex dynamics and the
  {F}ulde-{F}errell-{L}arkin-{O}vchinnikov state in a magnetic-field-induced
  organic superconductor},}\ }\href {\doibase 10.1103/PhysRevLett.97.157001}
  {\bibfield  {journal} {\bibinfo  {journal} {Phys. Rev. Lett.}\ }\textbf
  {\bibinfo {volume} {97}},\ \bibinfo {pages} {157001} (\bibinfo {year}
  {2006})}\BibitemShut {NoStop}%
\bibitem [{\citenamefont {Mayaffre}\ \emph {et~al.}(2014)\citenamefont
  {Mayaffre}, \citenamefont {Kr{\"a}mer}, \citenamefont {Horvati{\'c}},
  \citenamefont {Berthier}, \citenamefont {Miyagawa}, \citenamefont {Kanoda},\
  and\ \citenamefont {Mitrovi{\'c}}}]{Mayaffre2014}%
  \BibitemOpen
  \bibfield  {author} {\bibinfo {author} {\bibfnamefont {H.}~\bibnamefont
  {Mayaffre}}, \bibinfo {author} {\bibfnamefont {S.}~\bibnamefont
  {Kr{\"a}mer}}, \bibinfo {author} {\bibfnamefont {M.}~\bibnamefont
  {Horvati{\'c}}}, \bibinfo {author} {\bibfnamefont {C.}~\bibnamefont
  {Berthier}}, \bibinfo {author} {\bibfnamefont {K.}~\bibnamefont {Miyagawa}},
  \bibinfo {author} {\bibfnamefont {K.}~\bibnamefont {Kanoda}}, \ and\ \bibinfo
  {author} {\bibfnamefont {V.~F.}\ \bibnamefont {Mitrovi{\'c}}},\ }\bibfield
  {title} {\enquote {\bibinfo {title} {Evidence of {A}ndreev bound states as a
  hallmark of the {FFLO} phase in $\kappa$-({BEDT-TTF}$)_2${C}u({NCS}$)_2$},}\
  }\href {\doibase 10.1038/nphys3121} {\bibfield  {journal} {\bibinfo
  {journal} {Nature Physics}\ }\textbf {\bibinfo {volume} {10}},\ \bibinfo
  {pages} {928--932} (\bibinfo {year} {2014})}\BibitemShut {NoStop}%
\bibitem [{\citenamefont {Stein}\ \emph {et~al.}(2014)\citenamefont {Stein},
  \citenamefont {Sedrakian}, \citenamefont {Huang},\ and\ \citenamefont
  {Clark}}]{Stein2014}%
  \BibitemOpen
  \bibfield  {author} {\bibinfo {author} {\bibfnamefont {M.}~\bibnamefont
  {Stein}}, \bibinfo {author} {\bibfnamefont {A.}~\bibnamefont {Sedrakian}},
  \bibinfo {author} {\bibfnamefont {X.-G.}\ \bibnamefont {Huang}}, \ and\
  \bibinfo {author} {\bibfnamefont {J.~W.}\ \bibnamefont {Clark}},\ }\bibfield
  {title} {\enquote {\bibinfo {title} {{BCS}-{BEC} crossovers and
  unconventional phases in dilute nuclear matter},}\ }\href {\doibase
  10.1103/PhysRevC.90.065804} {\bibfield  {journal} {\bibinfo  {journal} {Phys.
  Rev. C}\ }\textbf {\bibinfo {volume} {90}},\ \bibinfo {pages} {065804}
  (\bibinfo {year} {2014})}\BibitemShut {NoStop}%
\bibitem [{\citenamefont {Alford}\ \emph {et~al.}(2008)\citenamefont {Alford},
  \citenamefont {Schmitt}, \citenamefont {Rajagopal},\ and\ \citenamefont
  {Sch\"afer}}]{Alford2008}%
  \BibitemOpen
  \bibfield  {author} {\bibinfo {author} {\bibfnamefont {M.~G.}\ \bibnamefont
  {Alford}}, \bibinfo {author} {\bibfnamefont {A.}~\bibnamefont {Schmitt}},
  \bibinfo {author} {\bibfnamefont {K.}~\bibnamefont {Rajagopal}}, \ and\
  \bibinfo {author} {\bibfnamefont {T.}~\bibnamefont {Sch\"afer}},\ }\bibfield
  {title} {\enquote {\bibinfo {title} {Color superconductivity in dense quark
  matter},}\ }\href {\doibase 10.1103/RevModPhys.80.1455} {\bibfield  {journal}
  {\bibinfo  {journal} {Rev. Mod. Phys.}\ }\textbf {\bibinfo {volume} {80}},\
  \bibinfo {pages} {1455--1515} (\bibinfo {year} {2008})}\BibitemShut {NoStop}%
\bibitem [{\citenamefont {Sedrakian}\ and\ \citenamefont
  {Rischke}(2009)}]{Sedrakian2009}%
  \BibitemOpen
  \bibfield  {author} {\bibinfo {author} {\bibfnamefont {A.}~\bibnamefont
  {Sedrakian}}\ and\ \bibinfo {author} {\bibfnamefont {D.~H.}\ \bibnamefont
  {Rischke}},\ }\bibfield  {title} {\enquote {\bibinfo {title} {Phase diagram
  of chiral quark matter: From weakly to strongly coupled {F}ulde-{F}errell
  phase},}\ }\href {\doibase 10.1103/PhysRevD.80.074022} {\bibfield  {journal}
  {\bibinfo  {journal} {Phys. Rev. D}\ }\textbf {\bibinfo {volume} {80}},\
  \bibinfo {pages} {074022} (\bibinfo {year} {2009})}\BibitemShut {NoStop}%
\bibitem [{\citenamefont {Huang}\ and\ \citenamefont
  {Sedrakian}(2010)}]{Huang2010}%
  \BibitemOpen
  \bibfield  {author} {\bibinfo {author} {\bibfnamefont {X.-G.}\ \bibnamefont
  {Huang}}\ and\ \bibinfo {author} {\bibfnamefont {A.}~\bibnamefont
  {Sedrakian}},\ }\bibfield  {title} {\enquote {\bibinfo {title} {Phase diagram
  of chiral quark matter: Color and electrically neutral {F}ulde-{F}errell
  phase},}\ }\href {\doibase 10.1103/PhysRevD.82.045029} {\bibfield  {journal}
  {\bibinfo  {journal} {Phys. Rev. D}\ }\textbf {\bibinfo {volume} {82}},\
  \bibinfo {pages} {045029} (\bibinfo {year} {2010})}\BibitemShut {NoStop}%
\bibitem [{\citenamefont {Wille}\ \emph {et~al.}(2008)\citenamefont {Wille},
  \citenamefont {Spiegelhalder}, \citenamefont {Kerner}, \citenamefont {Naik},
  \citenamefont {Trenkwalder}, \citenamefont {Hendl}, \citenamefont {Schreck},
  \citenamefont {Grimm}, \citenamefont {Tiecke}, \citenamefont {Walraven},
  \citenamefont {Kokkelmans}, \citenamefont {Tiesinga},\ and\ \citenamefont
  {Julienne}}]{Wille2008}%
  \BibitemOpen
  \bibfield  {author} {\bibinfo {author} {\bibfnamefont {E.}~\bibnamefont
  {Wille}}, \bibinfo {author} {\bibfnamefont {F.~M.}\ \bibnamefont
  {Spiegelhalder}}, \bibinfo {author} {\bibfnamefont {G.}~\bibnamefont
  {Kerner}}, \bibinfo {author} {\bibfnamefont {D.}~\bibnamefont {Naik}},
  \bibinfo {author} {\bibfnamefont {A.}~\bibnamefont {Trenkwalder}}, \bibinfo
  {author} {\bibfnamefont {G.}~\bibnamefont {Hendl}}, \bibinfo {author}
  {\bibfnamefont {F.}~\bibnamefont {Schreck}}, \bibinfo {author} {\bibfnamefont
  {R.}~\bibnamefont {Grimm}}, \bibinfo {author} {\bibfnamefont {T.~G.}\
  \bibnamefont {Tiecke}}, \bibinfo {author} {\bibfnamefont {J.~T.~M.}\
  \bibnamefont {Walraven}}, \bibinfo {author} {\bibfnamefont {S.~J. J. M.~F.}\
  \bibnamefont {Kokkelmans}}, \bibinfo {author} {\bibfnamefont
  {E.}~\bibnamefont {Tiesinga}}, \ and\ \bibinfo {author} {\bibfnamefont
  {P.~S.}\ \bibnamefont {Julienne}},\ }\bibfield  {title} {\enquote {\bibinfo
  {title} {Exploring an ultracold {F}ermi-{F}ermi mixture: Interspecies
  {F}eshbach resonances and scattering properties of $^{6}\mathrm{Li}$ and
  $^{40}\mathrm{K}$},}\ }\href {\doibase 10.1103/PhysRevLett.100.053201}
  {\bibfield  {journal} {\bibinfo  {journal} {Phys. Rev. Lett.}\ }\textbf
  {\bibinfo {volume} {100}},\ \bibinfo {pages} {053201} (\bibinfo {year}
  {2008})}\BibitemShut {NoStop}%
\bibitem [{\citenamefont {Voigt}\ \emph {et~al.}(2009)\citenamefont {Voigt},
  \citenamefont {Taglieber}, \citenamefont {Costa}, \citenamefont {Aoki},
  \citenamefont {Wieser}, \citenamefont {H\"ansch},\ and\ \citenamefont
  {Dieckmann}}]{Voigt2009}%
  \BibitemOpen
  \bibfield  {author} {\bibinfo {author} {\bibfnamefont {A.-C.}\ \bibnamefont
  {Voigt}}, \bibinfo {author} {\bibfnamefont {M.}~\bibnamefont {Taglieber}},
  \bibinfo {author} {\bibfnamefont {L.}~\bibnamefont {Costa}}, \bibinfo
  {author} {\bibfnamefont {T.}~\bibnamefont {Aoki}}, \bibinfo {author}
  {\bibfnamefont {W.}~\bibnamefont {Wieser}}, \bibinfo {author} {\bibfnamefont
  {T.~W.}\ \bibnamefont {H\"ansch}}, \ and\ \bibinfo {author} {\bibfnamefont
  {K.}~\bibnamefont {Dieckmann}},\ }\bibfield  {title} {\enquote {\bibinfo
  {title} {Ultracold heteronuclear {F}ermi-{F}ermi molecules},}\ }\href
  {\doibase 10.1103/PhysRevLett.102.020405} {\bibfield  {journal} {\bibinfo
  {journal} {Phys. Rev. Lett.}\ }\textbf {\bibinfo {volume} {102}},\ \bibinfo
  {pages} {020405} (\bibinfo {year} {2009})}\BibitemShut {NoStop}%
\bibitem [{\citenamefont {Jag}\ \emph {et~al.}(2014)\citenamefont {Jag},
  \citenamefont {Zaccanti}, \citenamefont {Cetina}, \citenamefont {Lous},
  \citenamefont {Schreck}, \citenamefont {Grimm}, \citenamefont {Petrov},\ and\
  \citenamefont {Levinsen}}]{Jag2014}%
  \BibitemOpen
  \bibfield  {author} {\bibinfo {author} {\bibfnamefont {M.}~\bibnamefont
  {Jag}}, \bibinfo {author} {\bibfnamefont {M.}~\bibnamefont {Zaccanti}},
  \bibinfo {author} {\bibfnamefont {M.}~\bibnamefont {Cetina}}, \bibinfo
  {author} {\bibfnamefont {R.~S.}\ \bibnamefont {Lous}}, \bibinfo {author}
  {\bibfnamefont {F.}~\bibnamefont {Schreck}}, \bibinfo {author} {\bibfnamefont
  {R.}~\bibnamefont {Grimm}}, \bibinfo {author} {\bibfnamefont {D.~S.}\
  \bibnamefont {Petrov}}, \ and\ \bibinfo {author} {\bibfnamefont
  {J.}~\bibnamefont {Levinsen}},\ }\bibfield  {title} {\enquote {\bibinfo
  {title} {Observation of a strong atom-dimer attraction in a mass-imbalanced
  {F}ermi-{F}ermi mixture},}\ }\href {\doibase 10.1103/PhysRevLett.112.075302}
  {\bibfield  {journal} {\bibinfo  {journal} {Phys. Rev. Lett.}\ }\textbf
  {\bibinfo {volume} {112}},\ \bibinfo {pages} {075302} (\bibinfo {year}
  {2014})}\BibitemShut {NoStop}%
\bibitem [{\citenamefont {Trenkwalder}\ \emph {et~al.}(2011)\citenamefont
  {Trenkwalder}, \citenamefont {Kohstall}, \citenamefont {Zaccanti},
  \citenamefont {Naik}, \citenamefont {Sidorov}, \citenamefont {Schreck},\ and\
  \citenamefont {Grimm}}]{Trenkwalder2014}%
  \BibitemOpen
  \bibfield  {author} {\bibinfo {author} {\bibfnamefont {A.}~\bibnamefont
  {Trenkwalder}}, \bibinfo {author} {\bibfnamefont {C.}~\bibnamefont
  {Kohstall}}, \bibinfo {author} {\bibfnamefont {M.}~\bibnamefont {Zaccanti}},
  \bibinfo {author} {\bibfnamefont {D.}~\bibnamefont {Naik}}, \bibinfo {author}
  {\bibfnamefont {A.~I.}\ \bibnamefont {Sidorov}}, \bibinfo {author}
  {\bibfnamefont {F.}~\bibnamefont {Schreck}}, \ and\ \bibinfo {author}
  {\bibfnamefont {R.}~\bibnamefont {Grimm}},\ }\bibfield  {title} {\enquote
  {\bibinfo {title} {Hydrodynamic expansion of a strongly interacting
  {F}ermi-{F}ermi mixture},}\ }\href {\doibase 10.1103/PhysRevLett.106.115304}
  {\bibfield  {journal} {\bibinfo  {journal} {Phys. Rev. Lett.}\ }\textbf
  {\bibinfo {volume} {106}},\ \bibinfo {pages} {115304} (\bibinfo {year}
  {2011})}\BibitemShut {NoStop}%
\bibitem [{\citenamefont {Gubbels}\ \emph {et~al.}(2009)\citenamefont
  {Gubbels}, \citenamefont {Baarsma},\ and\ \citenamefont
  {Stoof}}]{Gubbels2009}%
  \BibitemOpen
  \bibfield  {author} {\bibinfo {author} {\bibfnamefont {K.~B.}\ \bibnamefont
  {Gubbels}}, \bibinfo {author} {\bibfnamefont {J.~E.}\ \bibnamefont
  {Baarsma}}, \ and\ \bibinfo {author} {\bibfnamefont {H.~T.~C.}\ \bibnamefont
  {Stoof}},\ }\bibfield  {title} {\enquote {\bibinfo {title} {Lifshitz point in
  the phase diagram of resonantly interacting
  $^{6}\mathrm{Li}\mathrm{\text{-}}^{40}\mathrm{K}$ mixtures},}\ }\href
  {\doibase 10.1103/PhysRevLett.103.195301} {\bibfield  {journal} {\bibinfo
  {journal} {Phys. Rev. Lett.}\ }\textbf {\bibinfo {volume} {103}},\ \bibinfo
  {pages} {195301} (\bibinfo {year} {2009})}\BibitemShut {NoStop}%
\bibitem [{\citenamefont {Baarsma}\ \emph {et~al.}(2010)\citenamefont
  {Baarsma}, \citenamefont {Gubbels},\ and\ \citenamefont
  {Stoof}}]{Baarsma2010}%
  \BibitemOpen
  \bibfield  {author} {\bibinfo {author} {\bibfnamefont {J.~E.}\ \bibnamefont
  {Baarsma}}, \bibinfo {author} {\bibfnamefont {K.~B.}\ \bibnamefont
  {Gubbels}}, \ and\ \bibinfo {author} {\bibfnamefont {H.~T.~C.}\ \bibnamefont
  {Stoof}},\ }\bibfield  {title} {\enquote {\bibinfo {title} {Population and
  mass imbalance in atomic {F}ermi gases},}\ }\href {\doibase
  10.1103/PhysRevA.82.013624} {\bibfield  {journal} {\bibinfo  {journal} {Phys.
  Rev. A}\ }\textbf {\bibinfo {volume} {82}},\ \bibinfo {pages} {013624}
  (\bibinfo {year} {2010})}\BibitemShut {NoStop}%
\bibitem [{\citenamefont {Baarsma}\ and\ \citenamefont
  {Stoof}(2013)}]{Baarsma2013}%
  \BibitemOpen
  \bibfield  {author} {\bibinfo {author} {\bibfnamefont {J.~E.}\ \bibnamefont
  {Baarsma}}\ and\ \bibinfo {author} {\bibfnamefont {H.~T.~C.}\ \bibnamefont
  {Stoof}},\ }\bibfield  {title} {\enquote {\bibinfo {title} {Inhomogeneous
  superfluid phases in ${}^{6}${L}i-${}^{40}${K} mixtures at unitarity},}\
  }\href {\doibase 10.1103/PhysRevA.87.063612} {\bibfield  {journal} {\bibinfo
  {journal} {Phys. Rev. A}\ }\textbf {\bibinfo {volume} {87}},\ \bibinfo
  {pages} {063612} (\bibinfo {year} {2013})}\BibitemShut {NoStop}%
\bibitem [{\citenamefont {Wang}\ \emph {et~al.}(2013)\citenamefont {Wang},
  \citenamefont {Guo},\ and\ \citenamefont {Chen}}]{Wang2013}%
  \BibitemOpen
  \bibfield  {author} {\bibinfo {author} {\bibfnamefont {J.}~\bibnamefont
  {Wang}}, \bibinfo {author} {\bibfnamefont {H.}~\bibnamefont {Guo}}, \ and\
  \bibinfo {author} {\bibfnamefont {Q.}~\bibnamefont {Chen}},\ }\bibfield
  {title} {\enquote {\bibinfo {title} {Exotic phase separation and phase
  diagrams of a {F}ermi-{F}ermi mixture in a trap at finite temperature},}\
  }\href {\doibase 10.1103/PhysRevA.87.041601} {\bibfield  {journal} {\bibinfo
  {journal} {Phys. Rev. A}\ }\textbf {\bibinfo {volume} {87}},\ \bibinfo
  {pages} {041601} (\bibinfo {year} {2013})}\BibitemShut {NoStop}%
\bibitem [{\citenamefont {Pahl}\ and\ \citenamefont {Koinov}(2014)}]{Pahl2014}%
  \BibitemOpen
  \bibfield  {author} {\bibinfo {author} {\bibfnamefont {S.}~\bibnamefont
  {Pahl}}\ and\ \bibinfo {author} {\bibfnamefont {Z.}~\bibnamefont {Koinov}},\
  }\bibfield  {title} {\enquote {\bibinfo {title} {Phase diagram of a
  ${}^{6}\mathrm{Li}$-${}^{40}\mathrm{K}$ mixture in a square lattice},}\
  }\href {\doibase 10.1007/s10909-014-1166-9} {\bibfield  {journal} {\bibinfo
  {journal} {Journal of Low Temperature Physics}\ }\textbf {\bibinfo {volume}
  {176}},\ \bibinfo {pages} {113--121} (\bibinfo {year} {2014})}\BibitemShut
  {NoStop}%
\bibitem [{\citenamefont {Verstraete}\ and\ \citenamefont
  {Cirac}(2010)}]{Verstraete2010}%
  \BibitemOpen
  \bibfield  {author} {\bibinfo {author} {\bibfnamefont {F.}~\bibnamefont
  {Verstraete}}\ and\ \bibinfo {author} {\bibfnamefont {J.~I.}\ \bibnamefont
  {Cirac}},\ }\bibfield  {title} {\enquote {\bibinfo {title} {Continuous matrix
  product states for quantum fields},}\ }\href {\doibase
  10.1103/PhysRevLett.104.190405} {\bibfield  {journal} {\bibinfo  {journal}
  {Phys. Rev. Lett.}\ }\textbf {\bibinfo {volume} {104}},\ \bibinfo {pages}
  {190405} (\bibinfo {year} {2010})}\BibitemShut {NoStop}%
\bibitem [{\citenamefont {Haegeman}\ \emph
  {et~al.}(2013{\natexlab{a}})\citenamefont {Haegeman}, \citenamefont {Cirac},
  \citenamefont {Osborne},\ and\ \citenamefont {Verstraete}}]{Haegeman2013a}%
  \BibitemOpen
  \bibfield  {author} {\bibinfo {author} {\bibfnamefont {J.}~\bibnamefont
  {Haegeman}}, \bibinfo {author} {\bibfnamefont {J.~I.}\ \bibnamefont {Cirac}},
  \bibinfo {author} {\bibfnamefont {T.~J.}\ \bibnamefont {Osborne}}, \ and\
  \bibinfo {author} {\bibfnamefont {F.}~\bibnamefont {Verstraete}},\ }\bibfield
   {title} {\enquote {\bibinfo {title} {Calculus of continuous matrix product
  states},}\ }\href {\doibase 10.1103/PhysRevB.88.085118} {\bibfield  {journal}
  {\bibinfo  {journal} {Phys. Rev. B}\ }\textbf {\bibinfo {volume} {88}},\
  \bibinfo {pages} {085118} (\bibinfo {year} {2013}{\natexlab{a}})}\BibitemShut
  {NoStop}%
\bibitem [{\citenamefont {Affleck}\ \emph {et~al.}(1987)\citenamefont
  {Affleck}, \citenamefont {Kennedy}, \citenamefont {Lieb},\ and\ \citenamefont
  {Tasaki}}]{Affleck1987}%
  \BibitemOpen
  \bibfield  {author} {\bibinfo {author} {\bibfnamefont {I.}~\bibnamefont
  {Affleck}}, \bibinfo {author} {\bibfnamefont {T.}~\bibnamefont {Kennedy}},
  \bibinfo {author} {\bibfnamefont {E.~H.}\ \bibnamefont {Lieb}}, \ and\
  \bibinfo {author} {\bibfnamefont {H.}~\bibnamefont {Tasaki}},\ }\bibfield
  {title} {\enquote {\bibinfo {title} {Rigorous results on valence-bond ground
  states in antiferromagnets},}\ }\href {\doibase 10.1103/PhysRevLett.59.799}
  {\bibfield  {journal} {\bibinfo  {journal} {Phys. Rev. Lett.}\ }\textbf
  {\bibinfo {volume} {59}},\ \bibinfo {pages} {799--802} (\bibinfo {year}
  {1987})}\BibitemShut {NoStop}%
\bibitem [{\citenamefont {\"Ostlund}\ and\ \citenamefont
  {Rommer}(1995)}]{Ostlund1995}%
  \BibitemOpen
  \bibfield  {author} {\bibinfo {author} {\bibfnamefont {S.}~\bibnamefont
  {\"Ostlund}}\ and\ \bibinfo {author} {\bibfnamefont {S.}~\bibnamefont
  {Rommer}},\ }\bibfield  {title} {\enquote {\bibinfo {title} {Thermodynamic
  limit of density matrix renormalization},}\ }\href {\doibase
  10.1103/PhysRevLett.75.3537} {\bibfield  {journal} {\bibinfo  {journal}
  {Phys. Rev. Lett.}\ }\textbf {\bibinfo {volume} {75}},\ \bibinfo {pages}
  {3537--3540} (\bibinfo {year} {1995})}\BibitemShut {NoStop}%
\bibitem [{\citenamefont {Schollw\"ock}(2011)}]{Schollwock2011}%
  \BibitemOpen
  \bibfield  {author} {\bibinfo {author} {\bibfnamefont {U.}~\bibnamefont
  {Schollw\"ock}},\ }\bibfield  {title} {\enquote {\bibinfo {title} {The
  density-matrix renormalization group in the age of matrix product states},}\
  }\href {\doibase 10.1016/j.aop.2010.09.012} {\bibfield  {journal} {\bibinfo
  {journal} {Annals of Physics}\ }\textbf {\bibinfo {volume} {326}},\ \bibinfo
  {pages} {96 -- 192} (\bibinfo {year} {2011})}\BibitemShut {NoStop}%
\bibitem [{\citenamefont {White}(1992)}]{White1992}%
  \BibitemOpen
  \bibfield  {author} {\bibinfo {author} {\bibfnamefont {S.~R.}\ \bibnamefont
  {White}},\ }\bibfield  {title} {\enquote {\bibinfo {title} {Density matrix
  formulation for quantum renormalization groups},}\ }\href {\doibase
  10.1103/PhysRevLett.69.2863} {\bibfield  {journal} {\bibinfo  {journal}
  {Phys. Rev. Lett.}\ }\textbf {\bibinfo {volume} {69}},\ \bibinfo {pages}
  {2863--2866} (\bibinfo {year} {1992})}\BibitemShut {NoStop}%
\bibitem [{\citenamefont {Schollw\"ock}(2005)}]{Schollwock2005}%
  \BibitemOpen
  \bibfield  {author} {\bibinfo {author} {\bibfnamefont {U.}~\bibnamefont
  {Schollw\"ock}},\ }\bibfield  {title} {\enquote {\bibinfo {title} {The
  density-matrix renormalization group},}\ }\href {\doibase
  10.1103/RevModPhys.77.259} {\bibfield  {journal} {\bibinfo  {journal} {Rev.
  Mod. Phys.}\ }\textbf {\bibinfo {volume} {77}},\ \bibinfo {pages} {259--315}
  (\bibinfo {year} {2005})}\BibitemShut {NoStop}%
\bibitem [{\citenamefont {Rinc\'on}\ \emph {et~al.}(2015)\citenamefont
  {Rinc\'on}, \citenamefont {Ganahl},\ and\ \citenamefont
  {Vidal}}]{Rincon2015}%
  \BibitemOpen
  \bibfield  {author} {\bibinfo {author} {\bibfnamefont {J.}~\bibnamefont
  {Rinc\'on}}, \bibinfo {author} {\bibfnamefont {M.}~\bibnamefont {Ganahl}}, \
  and\ \bibinfo {author} {\bibfnamefont {G.}~\bibnamefont {Vidal}},\ }\bibfield
   {title} {\enquote {\bibinfo {title} {Lieb-{L}iniger model with exponentially
  decaying interactions: A continuous matrix product state study},}\ }\href
  {\doibase 10.1103/PhysRevB.92.115107} {\bibfield  {journal} {\bibinfo
  {journal} {Phys. Rev. B}\ }\textbf {\bibinfo {volume} {92}},\ \bibinfo
  {pages} {115107} (\bibinfo {year} {2015})}\BibitemShut {NoStop}%
\bibitem [{\citenamefont {Quijandr\'{\i}a}\ \emph {et~al.}(2014)\citenamefont
  {Quijandr\'{\i}a}, \citenamefont {Garc\'{\i}a-Ripoll},\ and\ \citenamefont
  {Zueco}}]{Quijandria2014}%
  \BibitemOpen
  \bibfield  {author} {\bibinfo {author} {\bibfnamefont {F.}~\bibnamefont
  {Quijandr\'{\i}a}}, \bibinfo {author} {\bibfnamefont {J.~J.}\ \bibnamefont
  {Garc\'{\i}a-Ripoll}}, \ and\ \bibinfo {author} {\bibfnamefont
  {D.}~\bibnamefont {Zueco}},\ }\bibfield  {title} {\enquote {\bibinfo {title}
  {Continuous matrix product states for coupled fields: Application to
  {L}uttinger liquids and quantum simulators},}\ }\href {\doibase
  10.1103/PhysRevB.90.235142} {\bibfield  {journal} {\bibinfo  {journal} {Phys.
  Rev. B}\ }\textbf {\bibinfo {volume} {90}},\ \bibinfo {pages} {235142}
  (\bibinfo {year} {2014})}\BibitemShut {NoStop}%
\bibitem [{\citenamefont {Quijandr\'{\i}a}\ and\ \citenamefont
  {Zueco}(2015)}]{Quijandria2015}%
  \BibitemOpen
  \bibfield  {author} {\bibinfo {author} {\bibfnamefont {F.}~\bibnamefont
  {Quijandr\'{\i}a}}\ and\ \bibinfo {author} {\bibfnamefont {D.}~\bibnamefont
  {Zueco}},\ }\bibfield  {title} {\enquote {\bibinfo {title}
  {Continuous-matrix-product-state solution for the mixing-demixing transition
  in one-dimensional quantum fields},}\ }\href {\doibase
  10.1103/PhysRevA.92.043629} {\bibfield  {journal} {\bibinfo  {journal} {Phys.
  Rev. A}\ }\textbf {\bibinfo {volume} {92}},\ \bibinfo {pages} {043629}
  (\bibinfo {year} {2015})}\BibitemShut {NoStop}%
\bibitem [{\citenamefont {Haegeman}\ \emph {et~al.}(2010)\citenamefont
  {Haegeman}, \citenamefont {Cirac}, \citenamefont {Osborne}, \citenamefont
  {Verschelde},\ and\ \citenamefont {Verstraete}}]{Haegeman2010}%
  \BibitemOpen
  \bibfield  {author} {\bibinfo {author} {\bibfnamefont {J.}~\bibnamefont
  {Haegeman}}, \bibinfo {author} {\bibfnamefont {J.~I.}\ \bibnamefont {Cirac}},
  \bibinfo {author} {\bibfnamefont {T.~J.}\ \bibnamefont {Osborne}}, \bibinfo
  {author} {\bibfnamefont {H.}~\bibnamefont {Verschelde}}, \ and\ \bibinfo
  {author} {\bibfnamefont {F.}~\bibnamefont {Verstraete}},\ }\bibfield  {title}
  {\enquote {\bibinfo {title} {Applying the variational principle to
  ($1+1$)-dimensional quantum field theories},}\ }\href {\doibase
  10.1103/PhysRevLett.105.251601} {\bibfield  {journal} {\bibinfo  {journal}
  {Phys. Rev. Lett.}\ }\textbf {\bibinfo {volume} {105}},\ \bibinfo {pages}
  {251601} (\bibinfo {year} {2010})}\BibitemShut {NoStop}%
\bibitem [{\citenamefont {Chung}\ \emph {et~al.}(2015)\citenamefont {Chung},
  \citenamefont {Sun},\ and\ \citenamefont {Bolech}}]{Chung2015}%
  \BibitemOpen
  \bibfield  {author} {\bibinfo {author} {\bibfnamefont {S.~S.}\ \bibnamefont
  {Chung}}, \bibinfo {author} {\bibfnamefont {K.}~\bibnamefont {Sun}}, \ and\
  \bibinfo {author} {\bibfnamefont {C.~J.}\ \bibnamefont {Bolech}},\ }\bibfield
   {title} {\enquote {\bibinfo {title} {Matrix product ansatz for {F}ermi
  fields in one dimension},}\ }\href {\doibase 10.1103/PhysRevB.91.121108}
  {\bibfield  {journal} {\bibinfo  {journal} {Phys. Rev. B}\ }\textbf {\bibinfo
  {volume} {91}},\ \bibinfo {pages} {121108} (\bibinfo {year}
  {2015})}\BibitemShut {NoStop}%
\bibitem [{\citenamefont {Draxler}\ \emph {et~al.}(2013)\citenamefont
  {Draxler}, \citenamefont {Haegeman}, \citenamefont {Osborne}, \citenamefont
  {Stojevic}, \citenamefont {Vanderstraeten},\ and\ \citenamefont
  {Verstraete}}]{Draxler2013}%
  \BibitemOpen
  \bibfield  {author} {\bibinfo {author} {\bibfnamefont {D.}~\bibnamefont
  {Draxler}}, \bibinfo {author} {\bibfnamefont {J.}~\bibnamefont {Haegeman}},
  \bibinfo {author} {\bibfnamefont {T.~J.}\ \bibnamefont {Osborne}}, \bibinfo
  {author} {\bibfnamefont {V.}~\bibnamefont {Stojevic}}, \bibinfo {author}
  {\bibfnamefont {L.}~\bibnamefont {Vanderstraeten}}, \ and\ \bibinfo {author}
  {\bibfnamefont {F.}~\bibnamefont {Verstraete}},\ }\bibfield  {title}
  {\enquote {\bibinfo {title} {Particles, holes, and solitons: A matrix product
  state approach},}\ }\href {\doibase 10.1103/PhysRevLett.111.020402}
  {\bibfield  {journal} {\bibinfo  {journal} {Phys. Rev. Lett.}\ }\textbf
  {\bibinfo {volume} {111}},\ \bibinfo {pages} {020402} (\bibinfo {year}
  {2013})}\BibitemShut {NoStop}%
\bibitem [{\citenamefont {Zaletel}\ and\ \citenamefont
  {Mong}(2012)}]{Zaletel2012}%
  \BibitemOpen
  \bibfield  {author} {\bibinfo {author} {\bibfnamefont {M.~P.}\ \bibnamefont
  {Zaletel}}\ and\ \bibinfo {author} {\bibfnamefont {R.~S.~K.}\ \bibnamefont
  {Mong}},\ }\bibfield  {title} {\enquote {\bibinfo {title} {Exact matrix
  product states for quantum {H}all wave functions},}\ }\href {\doibase
  10.1103/PhysRevB.86.245305} {\bibfield  {journal} {\bibinfo  {journal} {Phys.
  Rev. B}\ }\textbf {\bibinfo {volume} {86}},\ \bibinfo {pages} {245305}
  (\bibinfo {year} {2012})}\BibitemShut {NoStop}%
\bibitem [{\citenamefont {Cazalilla}\ \emph {et~al.}(2005)\citenamefont
  {Cazalilla}, \citenamefont {Ho},\ and\ \citenamefont
  {Giamarchi}}]{Cazalilla2005}%
  \BibitemOpen
  \bibfield  {author} {\bibinfo {author} {\bibfnamefont {M.~A.}\ \bibnamefont
  {Cazalilla}}, \bibinfo {author} {\bibfnamefont {A.~F.}\ \bibnamefont {Ho}}, \
  and\ \bibinfo {author} {\bibfnamefont {T.}~\bibnamefont {Giamarchi}},\
  }\bibfield  {title} {\enquote {\bibinfo {title} {Two-component {F}ermi gas on
  internal-state-dependent optical lattices},}\ }\href {\doibase
  10.1103/PhysRevLett.95.226402} {\bibfield  {journal} {\bibinfo  {journal}
  {Phys. Rev. Lett.}\ }\textbf {\bibinfo {volume} {95}},\ \bibinfo {pages}
  {226402} (\bibinfo {year} {2005})}\BibitemShut {NoStop}%
\bibitem [{\citenamefont {Parish}\ \emph {et~al.}(2007)\citenamefont {Parish},
  \citenamefont {Marchetti}, \citenamefont {Lamacraft},\ and\ \citenamefont
  {Simons}}]{Parish2007}%
  \BibitemOpen
  \bibfield  {author} {\bibinfo {author} {\bibfnamefont {M.~M.}\ \bibnamefont
  {Parish}}, \bibinfo {author} {\bibfnamefont {F.~M.}\ \bibnamefont
  {Marchetti}}, \bibinfo {author} {\bibfnamefont {A.}~\bibnamefont
  {Lamacraft}}, \ and\ \bibinfo {author} {\bibfnamefont {B.~D.}\ \bibnamefont
  {Simons}},\ }\bibfield  {title} {\enquote {\bibinfo {title} {Polarized
  {F}ermi condensates with unequal masses: Tuning the tricritical point},}\
  }\href {\doibase 10.1103/PhysRevLett.98.160402} {\bibfield  {journal}
  {\bibinfo  {journal} {Phys. Rev. Lett.}\ }\textbf {\bibinfo {volume} {98}},\
  \bibinfo {pages} {160402} (\bibinfo {year} {2007})}\BibitemShut {NoStop}%
\bibitem [{\citenamefont {Roscher}\ \emph {et~al.}(2014)\citenamefont
  {Roscher}, \citenamefont {Braun},\ and\ \citenamefont {Drut}}]{Roscher2014}%
  \BibitemOpen
  \bibfield  {author} {\bibinfo {author} {\bibfnamefont {D.}~\bibnamefont
  {Roscher}}, \bibinfo {author} {\bibfnamefont {J.}~\bibnamefont {Braun}}, \
  and\ \bibinfo {author} {\bibfnamefont {J.~E.}\ \bibnamefont {Drut}},\
  }\bibfield  {title} {\enquote {\bibinfo {title} {Inhomogeneous phases in
  one-dimensional mass- and spin-imbalanced {F}ermi gases},}\ }\href {\doibase
  10.1103/PhysRevA.89.063609} {\bibfield  {journal} {\bibinfo  {journal} {Phys.
  Rev. A}\ }\textbf {\bibinfo {volume} {89}},\ \bibinfo {pages} {063609}
  (\bibinfo {year} {2014})}\BibitemShut {NoStop}%
\bibitem [{\citenamefont {P\ifmmode~\mbox{\k{e}}\else \k{e}\fi{}cak}\ and\
  \citenamefont {Sowi\ifmmode~\acute{n}\else \'{n}\fi{}ski}(2016)}]{Pecak2016}%
  \BibitemOpen
  \bibfield  {author} {\bibinfo {author} {\bibfnamefont {D.}~\bibnamefont
  {P\ifmmode~\mbox{\k{e}}\else \k{e}\fi{}cak}}\ and\ \bibinfo {author}
  {\bibfnamefont {T.}~\bibnamefont {Sowi\ifmmode~\acute{n}\else
  \'{n}\fi{}ski}},\ }\bibfield  {title} {\enquote {\bibinfo {title} {Few
  strongly interacting ultracold fermions in one-dimensional traps of different
  shapes},}\ }\href {\doibase 10.1103/PhysRevA.94.042118} {\bibfield  {journal}
  {\bibinfo  {journal} {Phys. Rev. A}\ }\textbf {\bibinfo {volume} {94}},\
  \bibinfo {pages} {042118} (\bibinfo {year} {2016})}\BibitemShut {NoStop}%
\bibitem [{\citenamefont {Wang}\ \emph {et~al.}(2009)\citenamefont {Wang},
  \citenamefont {Chen},\ and\ \citenamefont {Das~Sarma}}]{Wang2009}%
  \BibitemOpen
  \bibfield  {author} {\bibinfo {author} {\bibfnamefont {B.}~\bibnamefont
  {Wang}}, \bibinfo {author} {\bibfnamefont {H.-D.}\ \bibnamefont {Chen}}, \
  and\ \bibinfo {author} {\bibfnamefont {S.}~\bibnamefont {Das~Sarma}},\
  }\bibfield  {title} {\enquote {\bibinfo {title} {Quantum phase diagram of
  fermion mixtures with population imbalance in one-dimensional optical
  lattices},}\ }\href {\doibase 10.1103/PhysRevA.79.051604} {\bibfield
  {journal} {\bibinfo  {journal} {Phys. Rev. A}\ }\textbf {\bibinfo {volume}
  {79}},\ \bibinfo {pages} {051604} (\bibinfo {year} {2009})}\BibitemShut
  {NoStop}%
\bibitem [{\citenamefont {Orso}\ \emph {et~al.}(2010)\citenamefont {Orso},
  \citenamefont {Burovski},\ and\ \citenamefont {Jolicoeur}}]{Orso2010}%
  \BibitemOpen
  \bibfield  {author} {\bibinfo {author} {\bibfnamefont {G.}~\bibnamefont
  {Orso}}, \bibinfo {author} {\bibfnamefont {E.}~\bibnamefont {Burovski}}, \
  and\ \bibinfo {author} {\bibfnamefont {T.}~\bibnamefont {Jolicoeur}},\
  }\bibfield  {title} {\enquote {\bibinfo {title} {Luttinger liquid of trimers
  in {F}ermi gases with unequal masses},}\ }\href {\doibase
  10.1103/PhysRevLett.104.065301} {\bibfield  {journal} {\bibinfo  {journal}
  {Phys. Rev. Lett.}\ }\textbf {\bibinfo {volume} {104}},\ \bibinfo {pages}
  {065301} (\bibinfo {year} {2010})}\BibitemShut {NoStop}%
\bibitem [{\citenamefont {Dalmonte}\ \emph {et~al.}(2012)\citenamefont
  {Dalmonte}, \citenamefont {Dieckmann}, \citenamefont {Roscilde},
  \citenamefont {Hartl}, \citenamefont {Feiguin}, \citenamefont
  {Schollw\"ock},\ and\ \citenamefont {Heidrich-Meisner}}]{Dalmonte2012}%
  \BibitemOpen
  \bibfield  {author} {\bibinfo {author} {\bibfnamefont {M.}~\bibnamefont
  {Dalmonte}}, \bibinfo {author} {\bibfnamefont {K.}~\bibnamefont {Dieckmann}},
  \bibinfo {author} {\bibfnamefont {T.}~\bibnamefont {Roscilde}}, \bibinfo
  {author} {\bibfnamefont {C.}~\bibnamefont {Hartl}}, \bibinfo {author}
  {\bibfnamefont {A.~E.}\ \bibnamefont {Feiguin}}, \bibinfo {author}
  {\bibfnamefont {U.}~\bibnamefont {Schollw\"ock}}, \ and\ \bibinfo {author}
  {\bibfnamefont {F.}~\bibnamefont {Heidrich-Meisner}},\ }\bibfield  {title}
  {\enquote {\bibinfo {title} {Dimer, trimer, and
  {F}ulde-{F}errell-{L}arkin-{O}vchinnikov liquids in mass- and spin-imbalanced
  trapped binary mixtures in one dimension},}\ }\href {\doibase
  10.1103/PhysRevA.85.063608} {\bibfield  {journal} {\bibinfo  {journal} {Phys.
  Rev. A}\ }\textbf {\bibinfo {volume} {85}},\ \bibinfo {pages} {063608}
  (\bibinfo {year} {2012})}\BibitemShut {NoStop}%
\bibitem [{\citenamefont {Stoudenmire}\ \emph {et~al.}(2012)\citenamefont
  {Stoudenmire}, \citenamefont {Wagner}, \citenamefont {White},\ and\
  \citenamefont {Burke}}]{Stoudenmire2012}%
  \BibitemOpen
  \bibfield  {author} {\bibinfo {author} {\bibfnamefont {E.~M.}\ \bibnamefont
  {Stoudenmire}}, \bibinfo {author} {\bibfnamefont {L.~O.}\ \bibnamefont
  {Wagner}}, \bibinfo {author} {\bibfnamefont {S.~R.}\ \bibnamefont {White}}, \
  and\ \bibinfo {author} {\bibfnamefont {K.}~\bibnamefont {Burke}},\ }\bibfield
   {title} {\enquote {\bibinfo {title} {One-dimensional continuum electronic
  structure with the density-matrix renormalization group and its implications
  for density-functional theory},}\ }\href {\doibase
  10.1103/PhysRevLett.109.056402} {\bibfield  {journal} {\bibinfo  {journal}
  {Phys. Rev. Lett.}\ }\textbf {\bibinfo {volume} {109}},\ \bibinfo {pages}
  {056402} (\bibinfo {year} {2012})}\BibitemShut {NoStop}%
\bibitem [{\citenamefont {Dolfi}\ \emph {et~al.}(2012)\citenamefont {Dolfi},
  \citenamefont {Bauer}, \citenamefont {Troyer},\ and\ \citenamefont
  {Ristivojevic}}]{Dolfi2012}%
  \BibitemOpen
  \bibfield  {author} {\bibinfo {author} {\bibfnamefont {M.}~\bibnamefont
  {Dolfi}}, \bibinfo {author} {\bibfnamefont {B.}~\bibnamefont {Bauer}},
  \bibinfo {author} {\bibfnamefont {M.}~\bibnamefont {Troyer}}, \ and\ \bibinfo
  {author} {\bibfnamefont {Z.}~\bibnamefont {Ristivojevic}},\ }\bibfield
  {title} {\enquote {\bibinfo {title} {Multigrid algorithms for tensor network
  states},}\ }\href {\doibase 10.1103/PhysRevLett.109.020604} {\bibfield
  {journal} {\bibinfo  {journal} {Phys. Rev. Lett.}\ }\textbf {\bibinfo
  {volume} {109}},\ \bibinfo {pages} {020604} (\bibinfo {year}
  {2012})}\BibitemShut {NoStop}%
\bibitem [{\citenamefont {Yang}(2001)}]{Yang2001}%
  \BibitemOpen
  \bibfield  {author} {\bibinfo {author} {\bibfnamefont {K.}~\bibnamefont
  {Yang}},\ }\bibfield  {title} {\enquote {\bibinfo {title} {Inhomogeneous
  superconducting state in quasi-one-dimensional systems},}\ }\href {\doibase
  10.1103/PhysRevB.63.140511} {\bibfield  {journal} {\bibinfo  {journal} {Phys.
  Rev. B}\ }\textbf {\bibinfo {volume} {63}},\ \bibinfo {pages} {140511}
  (\bibinfo {year} {2001})}\BibitemShut {NoStop}%
\bibitem [{\citenamefont {Haegeman}\ \emph
  {et~al.}(2013{\natexlab{b}})\citenamefont {Haegeman}, \citenamefont
  {Osborne}, \citenamefont {Verschelde},\ and\ \citenamefont
  {Verstraete}}]{Haegeman2013c}%
  \BibitemOpen
  \bibfield  {author} {\bibinfo {author} {\bibfnamefont {J.}~\bibnamefont
  {Haegeman}}, \bibinfo {author} {\bibfnamefont {T.~J.}\ \bibnamefont
  {Osborne}}, \bibinfo {author} {\bibfnamefont {H.}~\bibnamefont {Verschelde}},
  \ and\ \bibinfo {author} {\bibfnamefont {F.}~\bibnamefont {Verstraete}},\
  }\bibfield  {title} {\enquote {\bibinfo {title} {Entanglement renormalization
  for quantum fields in real space},}\ }\href {\doibase
  10.1103/PhysRevLett.110.100402} {\bibfield  {journal} {\bibinfo  {journal}
  {Phys. Rev. Lett.}\ }\textbf {\bibinfo {volume} {110}},\ \bibinfo {pages}
  {100402} (\bibinfo {year} {2013}{\natexlab{b}})}\BibitemShut {NoStop}%
\bibitem [{\citenamefont {Evenbly}\ and\ \citenamefont
  {Vidal}(2013)}]{Evenbly2013}%
  \BibitemOpen
  \bibfield  {author} {\bibinfo {author} {\bibfnamefont {G.}~\bibnamefont
  {Evenbly}}\ and\ \bibinfo {author} {\bibfnamefont {G.}~\bibnamefont
  {Vidal}},\ }\bibfield  {title} {\enquote {\bibinfo {title} {Quantum
  criticality with the multi-scale entanglement renormalization ansatz},}\ }in\
  \href@noop {} {\emph {\bibinfo {booktitle} {Strongly Correlated Systems}}}\
  (\bibinfo  {publisher} {Springer},\ \bibinfo {year} {2013})\ pp.\ \bibinfo
  {pages} {99--130}\BibitemShut {NoStop}%
\bibitem [{\citenamefont {Moritz}\ \emph {et~al.}(2005)\citenamefont {Moritz},
  \citenamefont {St\"oferle}, \citenamefont {G\"unter}, \citenamefont
  {K\"ohl},\ and\ \citenamefont {Esslinger}}]{Moritz2005}%
  \BibitemOpen
  \bibfield  {author} {\bibinfo {author} {\bibfnamefont {H.}~\bibnamefont
  {Moritz}}, \bibinfo {author} {\bibfnamefont {T.}~\bibnamefont {St\"oferle}},
  \bibinfo {author} {\bibfnamefont {K.}~\bibnamefont {G\"unter}}, \bibinfo
  {author} {\bibfnamefont {M.}~\bibnamefont {K\"ohl}}, \ and\ \bibinfo {author}
  {\bibfnamefont {T.}~\bibnamefont {Esslinger}},\ }\bibfield  {title} {\enquote
  {\bibinfo {title} {Confinement induced molecules in a 1d {F}ermi gas},}\
  }\href {\doibase 10.1103/PhysRevLett.94.210401} {\bibfield  {journal}
  {\bibinfo  {journal} {Phys. Rev. Lett.}\ }\textbf {\bibinfo {volume} {94}},\
  \bibinfo {pages} {210401} (\bibinfo {year} {2005})}\BibitemShut {NoStop}%
\bibitem [{\citenamefont {Baksmaty}\ \emph {et~al.}(2011)\citenamefont
  {Baksmaty}, \citenamefont {Lu}, \citenamefont {Bolech},\ and\ \citenamefont
  {Pu}}]{Baksmaty2011}%
  \BibitemOpen
  \bibfield  {author} {\bibinfo {author} {\bibfnamefont {L.~O.}\ \bibnamefont
  {Baksmaty}}, \bibinfo {author} {\bibfnamefont {H.}~\bibnamefont {Lu}},
  \bibinfo {author} {\bibfnamefont {C.~J.}\ \bibnamefont {Bolech}}, \ and\
  \bibinfo {author} {\bibfnamefont {H.}~\bibnamefont {Pu}},\ }\bibfield
  {title} {\enquote {\bibinfo {title} {Concomitant modulated superfluidity in
  polarized {F}ermi gases},}\ }\href {\doibase 10.1103/PhysRevA.83.023604}
  {\bibfield  {journal} {\bibinfo  {journal} {Phys. Rev. A}\ }\textbf {\bibinfo
  {volume} {83}},\ \bibinfo {pages} {023604} (\bibinfo {year}
  {2011})}\BibitemShut {NoStop}%
\bibitem [{\citenamefont {Son}\ and\ \citenamefont
  {Stephanov}(2006)}]{Son2006}%
  \BibitemOpen
  \bibfield  {author} {\bibinfo {author} {\bibfnamefont {D.~T.}\ \bibnamefont
  {Son}}\ and\ \bibinfo {author} {\bibfnamefont {M.~A.}\ \bibnamefont
  {Stephanov}},\ }\bibfield  {title} {\enquote {\bibinfo {title} {Phase diagram
  of a cold polarized fermi gas},}\ }\href {\doibase
  10.1103/PhysRevA.74.013614} {\bibfield  {journal} {\bibinfo  {journal} {Phys.
  Rev. A}\ }\textbf {\bibinfo {volume} {74}},\ \bibinfo {pages} {013614}
  (\bibinfo {year} {2006})}\BibitemShut {NoStop}%
\bibitem [{\citenamefont {Bulgac}\ \emph {et~al.}(2006)\citenamefont {Bulgac},
  \citenamefont {Forbes},\ and\ \citenamefont {Schwenk}}]{Bulgac2006}%
  \BibitemOpen
  \bibfield  {author} {\bibinfo {author} {\bibfnamefont {A.}~\bibnamefont
  {Bulgac}}, \bibinfo {author} {\bibfnamefont {M.~M.}\ \bibnamefont {Forbes}},
  \ and\ \bibinfo {author} {\bibfnamefont {A.}~\bibnamefont {Schwenk}},\
  }\bibfield  {title} {\enquote {\bibinfo {title} {Induced $p$-wave
  superfluidity in asymmetric fermi gases},}\ }\href {\doibase
  10.1103/PhysRevLett.97.020402} {\bibfield  {journal} {\bibinfo  {journal}
  {Phys. Rev. Lett.}\ }\textbf {\bibinfo {volume} {97}},\ \bibinfo {pages}
  {020402} (\bibinfo {year} {2006})}\BibitemShut {NoStop}%
\bibitem [{\citenamefont {Mei}\ and\ \citenamefont {Bolech}(2017)}]{Mei2016}%
  \BibitemOpen
  \bibfield  {author} {\bibinfo {author} {\bibfnamefont {Z.}~\bibnamefont
  {Mei}}\ and\ \bibinfo {author} {\bibfnamefont {C.~J.}\ \bibnamefont
  {Bolech}},\ }\bibfield  {title} {\enquote {\bibinfo {title} {Derivation of
  matrix product states for the {H}eisenberg spin chain with open boundary
  conditions},}\ }\href {\doibase 10.1103/PhysRevE.95.032127} {\bibfield
  {journal} {\bibinfo  {journal} {Phys. Rev. E}\ }\textbf {\bibinfo {volume}
  {95}},\ \bibinfo {pages} {032127} (\bibinfo {year} {2017})}\BibitemShut
  {NoStop}%
\bibitem [{\citenamefont {Ganahl}\ \emph {et~al.}(2017)\citenamefont {Ganahl},
  \citenamefont {Rinc\'on},\ and\ \citenamefont {Vidal}}]{Ganahl2017}%
  \BibitemOpen
  \bibfield  {author} {\bibinfo {author} {\bibfnamefont {M.}~\bibnamefont
  {Ganahl}}, \bibinfo {author} {\bibfnamefont {J.}~\bibnamefont {Rinc\'on}}, \
  and\ \bibinfo {author} {\bibfnamefont {G.}~\bibnamefont {Vidal}},\ }\bibfield
   {title} {\enquote {\bibinfo {title} {Continuous matrix product states for
  quantum fields: An energy minimization algorithm},}\ }\href {\doibase
  10.1103/PhysRevLett.118.220402} {\bibfield  {journal} {\bibinfo  {journal}
  {Phys. Rev. Lett.}\ }\textbf {\bibinfo {volume} {118}},\ \bibinfo {pages}
  {220402} (\bibinfo {year} {2017})}\BibitemShut {NoStop}%
\bibitem [{Ohi()}]{OhioSupercomputerCenter1987}%
  \BibitemOpen
  \href@noop {} {\enquote {\bibinfo {title} {{O}hio {S}upercomputer
  {C}enter},}\ }\bibinfo {howpublished}
  {\url{http://osc.edu/ark:/19495/f5s1ph73}}\BibitemShut {NoStop}%
\end{thebibliography}
%

\end{document}